\documentclass[letter]{IEEEtran}
\IEEEoverridecommandlockouts
\usepackage{epsfig}
\usepackage{graphicx}
\usepackage{amsmath}
\usepackage{mathtools}
\usepackage{amssymb}
\usepackage{verbatim}
\usepackage{mathrsfs}
\usepackage{algorithm}
\usepackage{subfigure}
\usepackage[noend]{algpseudocode}
\usepackage{xcolor}% you could also use the color package
\usepackage{colortbl}
\usepackage{lipsum}
\usepackage{amsmath,amssymb}
\usepackage{amssymb}%
\usepackage{mathrsfs}%
\usepackage{amsthm}

\usepackage{bm}
\allowdisplaybreaks
\usepackage[noend]{algpseudocode}

\begin{document}
\title{Decentralized Federated Learning with Unreliable Communications}
\author{\IEEEauthorblockN{Hao Ye, Le Liang, and Geoffrey Ye Li}
\thanks{
H. Ye is with the School of Electrical and Computer Engineering, Georgia Institute of Technology. (email: yehao@gatech.edu)
L. Liang is with School of Information Science and Engineering, Southeast University, Nanjing 210096, China. (email: lliang@seu.edu.cn)
G. Y. Li is with the Department of Electrical and Electronic Engineering, Imperial College London. (email: geoffrey.li@imperial.ac.uk)
}
}
\maketitle
\begin{abstract}
    Decentralized federated learning, inherited from decentralized learning, enables the edge devices to collaborate on model training in a peer-to-peer manner without the assistance of a server. 
    However, existing decentralized learning frameworks usually assume perfect communication among devices, where they can reliably exchange messages, \emph{e.g.}, gradients or parameters.
    But the real-world communication networks are prone to packet loss and transmission errors.
    Transmission reliability comes with a price. 
    The commonly-used solution is to adopt a reliable transportation layer protocol, \emph{e.g.}, transmission control protocol (TCP), which however leads to significant communication overhead and reduces connectivity among devices that can be supported.
    For a communication network with a lightweight and unreliable communication protocol, user datagram protocol (UDP), we propose  a robust decentralized stochastic gradient descent (SGD) approach, called Soft-DSGD, to address the unreliability issue.
    Soft-DSGD updates the model parameters with partially received messages and optimizes the mixing weights according to the link reliability matrix of communication links. 
    We prove that the proposed decentralized training system, even with unreliable communications, can still achieve the same asymptotic convergence rate as vanilla decentralized SGD with perfect communications.
    Moreover, numerical results confirm the proposed approach can leverage all available unreliable communication links to speed up  convergence.
\end{abstract}

\section{Introduction}

Empowered by massive training data, deep learning is emerging as the major driving force behind breakthroughs in a wide range of areas, including computer vision, natural language processing, speech processing, etc. 
Inspired by these success stories, there is a growing trend to use deep learning to unleash the power of massive data generated at the edge network by such devices as smart phones, wearables, and sensors \cite{qin2020federated}.
Nevertheless, the limited communication bandwidth, together with the desire for user privacy, makes it infeasible to collect all distributed data and then train a model at a single machine.
As an alternative, training deep learning models at edge networks in a distributed fashion has gained much attention in, \emph{e.g.,} federated learning \cite{qin2020federated}.
It enables multiple devices to compute multiple devices compute the local mini-batch gradients of the loss function with their local datasets and only exchange model parameters or gradients over a communication network to ensure convergence to the global optimal solution, and protect the privacy of the local devices.

\begin{figure}
    \subfigure[]{\includegraphics[width=0.45\textwidth]{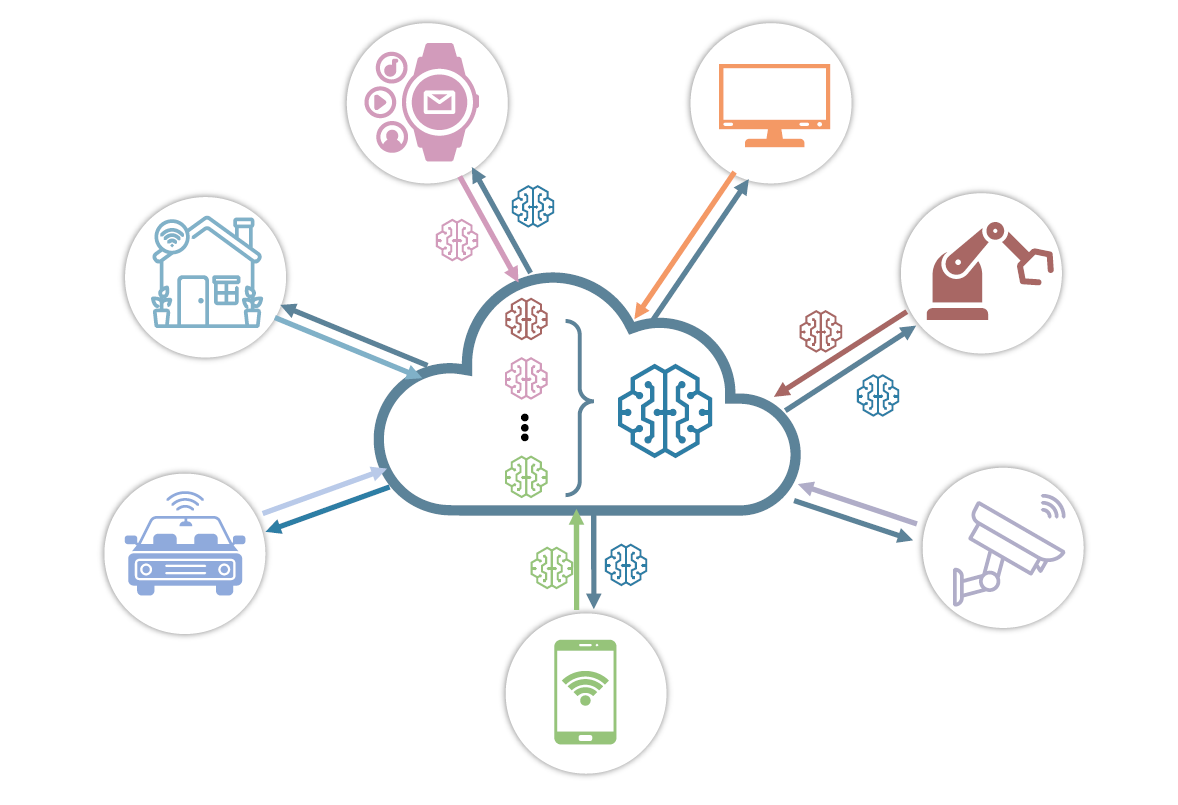}} 
    \subfigure[]{\includegraphics[width=0.45\textwidth]{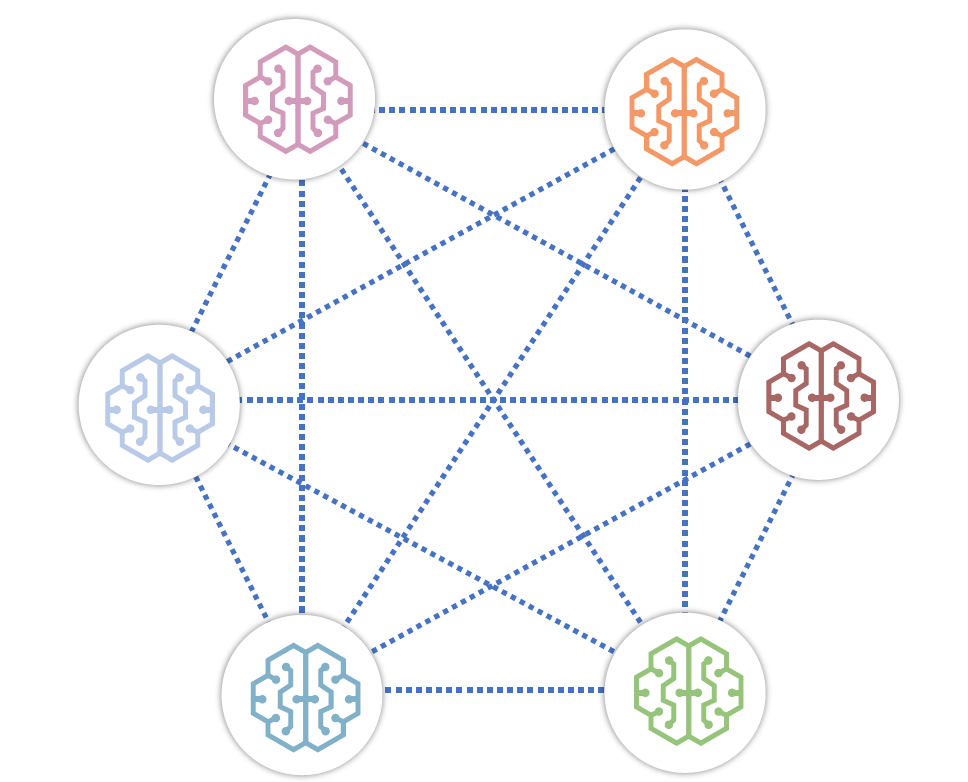}}
    \caption{ (a) Centralized federated learning. (b) Decentralized federated learning.} \label{fig:FL}
\end{figure}

Traditional federated learning is designed with a parameter-server architecture as shown in Fig. \ref{fig:FL} (a), where a centralized server orchestrates the training process.  
In each training iteration, the global model is transmitted to the participating edge devices and these devices compute a set of potential model updates based on local data. 
These updates are then sent to the central server and aggregated into a single global update. 
Since all the selected devices have to send updates to a single centralized server in each training iteration, the server becomes a communication bottleneck of the system, which makes it difficult to scale to a large number of devices.

In this paper, we focus on the decentralized training paradigm building on the device-to-device (D2D) network, which facilitates collaborative learning made possible by the ability to communicate among the edge devices without the assistance of the server.
As shown in Fig. \ref{fig:FL} (b), each device stores and optimizes the model parameters with its own data and exchanges messages with neighbours in a peer-to-peer manner to reach a consensus. Decentralized optimization has been investigated intensively since 1980s \cite{tsitsiklis1984problems}. 
The most widely-used decentralized algorithms include gradient and subgradient descent \cite{nedic2009distributed, yuan2016convergence}, alternating direction method of multipliers (ADMM) \cite{wei2012distributed, shi2014linear, zheng2018game}, and dual averaging \cite{duchi2011dual}.
Recently, decentralized implementations of SGD for training deep learning models have gained much attention \cite{lian2017can, lian2018asynchronous, assran2019stochastic,  wang2018cooperative, wang2019matcha}.
The decentralized stochastic gradient descent (SGD) algorithm has been investigated in \cite{lian2017can} for optimizing non-convex objectives (\emph{e.g.}, training deep neural networks) and proven to have the same asymptotic convergence rate as the centralized SGD.
This framework has been extended to deal with variations in communication paradigms. An asynchronous decentralized SGD framework has been developed in \cite{lian2018asynchronous}, where the devices communicate asynchronously to reduce the idle time. 
In \cite{assran2019stochastic}, gradient push-sum has been exploited to develop a decentralized SGD under a directed communication network.

A major bottleneck in achieving fast convergence is the limited communication resources, such as bandwidth and power.
To reduce the communication cost in decentralized training, prior works on building communication-efficient decentralized training systems have concentrated on compressing the communication messages via sparsification and quantization \cite{tang2018communication,tang2019doublesqueeze, koloskova2019decentralized},  skipping the communication rounds by performing a certain number of local updates \cite{wang2018cooperative, koloskova2020unified}, or communicating in an asynchronous manner \cite{lian2018asynchronous}.
% Decentralized SGD algorithms with compressed communication messages have been proposed to save the communication cost.
It has been shown in \cite{tang2018communication} that directly compressing the shared messages leads to non-vanishing quantization errors, which in turn causes failure to converge.
The first exact decentralized optimization method with message quantization has been developed in \cite{reisizadeh2019exact} for strongly convex objectives. 
For non-convex optimization, quantization approaches have been proposed  in \cite{tang2018communication}, where the differences of parameters of two consecutive steps are quantized and shared.
Moreover, DeepSqueeze in \cite{tang2019doublesqueeze} applies an error-compensation method to decentralized settings. 
ChocoSGD in \cite{koloskova2019decentralized}  lets devices estimate remote models with a local estimator, which supports arbitrary quantization by tuning the communication matrix.
Apart from compressing the communication messages, another direction to improve the communication efficiency is reducing the communication frequency.
The devices take several local SGD update steps before  a consensus update with other devices \cite{wang2018cooperative, koloskova2020unified}.

Nevertheless, most of the existing decentralized optimization methods require a reliable communication network among the devices while the real-world communication systems are prone to packet loss and transmission errors.
% In particular, with the development of federated learning, Internet of things (IoT), and edge computing, many machine learning models tend to be trained at the edge network and wireless communications will be the major communication type, where transmission errors are pervasive due to the uncontrollable environment.
The transmission errors are pervasive in federated learning due to the harsh wireless channels, which introduce noise,  fading, and interference.
The default solution is to use a reliable transportation layer communication protocol, \textit{e.g.,} transmission control protocol (TCP), where acknowledgment (ACK), retransmission, and time-out mechanisms are employed to detect and recover from transmission failures.
But this transmission reliability comes with a price, which usually incurs considerable communication overheads since the messages may be transmitted multiple times.
In addition, the number of neighbours that each device can communicate with is limited to ensure reliability, which also slows down the training speed.

% Both TCP and UDP are widely-used transportation layer protocols for transmitting data packets over a communication network.
% TCP is a connection-oriented protocol and requires handshakes to set up an end-to-end communication connection, where messages can be transmitted in both directions. 
% Hence, TCP has a considerable communication overhead, especially in unreliable communication networks, such as wireless systems.

% Compared to TCP, UDP is a lightweight message-based connectionless protocol with much less overhead and does not need to set up a dedicated connection for transmission. 
% Communication is achieved by transmitting packets in one direction from the source to the destination without verifying whether the recipients have received the packets.
% Due to the unreliability nature of UDP, it is mainly used for delivery best effort traffic, such as multimedia streaming, where occasional packet loss can be tolerated. 
% In this paper, we show that UDP can also be applied in decentralized training as the random packet loss can be tolerated with purposely designed frameworks.

In this paper,  a lightweight transmission protocol, user datagram protocol (UDP), is adopted to provide connectionless and unreliable packet delivery service instead of using these reliable but heavyweight communication protocols.
Each device communicates with the rest of the network via soft and unreliable communication links, where packet loss and transmission errors occur randomly.
A robust decentralized training algorithm, called Soft-DSGD, is developed to deal with the unreliable communications.
The devices update their model parameters with partially received messages and the mixing weights in the consensus updates are optimized according to the reliability matrix of different communication links. 
We prove that the proposed Soft-DSGD under the unreliable communication network achieves the same asymptotic convergence rate  as the vanilla decentralized SGD with perfect communications. 
In addition, numerical results confirm that Soft-DSGD can efficiently leverage all unreliable communication links that are available to accelerate convergence.

The rest of the paper is organized as follows. In Section \ref{sec:Related_Work}, the background information on the decentralized optimization is provided. In Section \ref{sec:Method}, the proposed Soft-DSGD for training deep learning models with unreliable communication networks is presented in detail. The theoretical convergence analysis of Soft-DSGD is shown in Section \ref{sec:Theory}. The simulation results are presented in Section \ref{sec:Exp} and the conclusions are drawn in Section \ref{sec:Conclusion}.

\section{Decentralized Optimization} \label{sec:Related_Work}
In this section, we briefly introduce the decentralized optimization, including the setting and the applications.

% There are few existing works investigating the distributed optimization methods with an unreliable communication network, where the communication links fail randomly.
% An exception is \cite{yu2019distributed}, which developed a robust distributed training framework for the parameter-server architecture. 
% In addition, this paper is also closely related to gossiping-based consensus algorithms, where the devices in a network seek to converge their states to the average of their initial states in a distributed manner \cite{boyd2006randomized, aysal2009broadcast}.  Especially, for the consensus problem with wireless networks, the time-varying communication topologies are commonly-used to model the noisy and unreliable communication links \cite{olfati2004consensus, kar2008sensor, kar2008distributed, tahbaz2008necessary} and optimal consensus weights have been investigated in order to obtain the best convergence rate \cite{aysal2009broadcast, jakovetic2010weight}. 

\subsection{Decentralized Optimization and Decentralized SGD} 
Here, we briefly introduce decentralized training. We consider the following decentralized optimization problem over a network of $N$ devices:  
\begin{equation*}
    f^* := \min_{\mathbf{x} \in \mathbb{R}^d} \left[f(\mathbf{x}) := \frac{1}{N} \sum_{i=1}^N f_i(\mathbf{x}) \right],
\end{equation*}
where each component $f_i: \mathbb{R}^d \rightarrow \mathbb{R}$ defines the local objective and is only known by the $i$-th device.
The network topology is represented by an undirected connected graph $\mathcal{G}$, where the devices can only communicate along the edges.
For deep learning model training, the local objective $f_i$ is given in a stochastic form,
\begin{equation*}
    f_i(\mathbf{x}) := \mathbb{E}_{\xi_i \sim \mathcal{D}_i} F_i(\mathbf{x}, \xi_i),
\end{equation*}
where $\mathcal{D}_i$ denotes the local data stored at device $i$, $\xi_i$ is a mini-batch data from $\mathcal{D}_i$, and $F_i(\mathbf{x}, \xi_i)$ is the local loss function with respect to $\xi_i$.

In the decentralized SGD framework \cite{lian2017can}, each device maintains its own local parameters, $\mathbf{x}_i \in \mathbb{R}^d$, and computes the local gradient, $\mathbf{g}_i := \nabla F_i(\mathbf{x}_i, \xi_i)$ based on $\xi_i$ sampled from the local dataset. After that, parameters are exchanged with the neighbouring devices via peer-to-peer communications.
Formally, the two steps in each training iteration are 
\begin{enumerate}
    \item \textbf{SGD update:} $\mathbf{x}_i^{(t+\frac{1}{2})} = \mathbf{x}_i^{(t)} - \gamma \mathbf{g}_i^{(t)}$, where $\gamma$ denotes the learning rate.
    \item \textbf{Consensus update:} $\mathbf{x}_i^{(t+1)} = \sum_{j=1}^{N} w_{i,j}\mathbf{x}_j^{(t+\frac{1}{2})}$, where $w_{i,j}$ is the $(i, j)$-th item of the mixing weights matrix $\mathbf{W} \in \mathbb{R}^{N\times N}$  and $w_{i,j} \neq 0$ only if the $i$-th device and the $j$-th device are neighbours in $\mathcal{G}$.  
\end{enumerate}
To ensure convergence,  $\mathbf{W}$ is often set to be symmetric and doubly stochastic, \textit{i.e.,} $\mathbf{W}^T = \mathbf{W}$ and $\mathbf{W} \mathbf{1}  = \mathbf{1}$, where $\mathbf{1}$ indicates an $N$-dimensional vector of all $1$’s.
The spectrum gap of $\mathbf{W}$ is also required to be strictly positive, \emph{i.e.}, $1 - \max\{ \| \lambda_2(\mathbf{W})\|, \|\lambda_N(\mathbf{W}) \|\} > 0$, where $\lambda_k(\mathbf{W})$ denotes the $k$-th largest eigenvalue of $\mathbf{W}$.

% The convergence of the decentralized SGD has been investigated before.

% \textbf{(Convergence of D-SGD \cite{lian2017can})} Under certain assumptions, the output of the D-PSGD admits the following inequality

\subsection{Motivating Applications} 
The recent increasing attention on the decentralized optimization is mainly driven by a wide range of applications where a network of devices need to cooperate to optimize the common objective. 
Three examples are included here and more applications of the decentralized optimization can be found in \cite{cao2012overview}.

\subsubsection{Decentralized estimation} 
The wireless sensor network and internet-of-things (IoT) are often utilized for monitoring and estimating uncertain environmental state $s$.  
Suppose there are $N$ devices and each device has a measurement $y_i$, which is modeled as a random variable with density $p_i(y_i|s)$.
Due to the noise, collaboration among the devices leads to a more robust estimation.
When the measurement dimension is large, \emph{e.g.,} images and videos, it becomes more efficient to perform decentralized estimation rather than centralized estimation. 
In this case, the maximum likelihood estimate of $s$ can obtained by solving
\begin{equation}
    \max_s \sum_{i=1}^{N} \log p_i(y_i|s),
\end{equation}
which can be addressed by the decentralized optimization framework with $f_i = p_i(y_i|s)$.

\subsubsection{Decentralized resource allocation}
The increasing user demand and number of devices impose critical challenges on the wireless resource management schemes, which aim at making the best use of the limited resources (e.g. bandwidth and power).
The objective of resource allocation can be to maximize the summation of a utility function of all communication links, \textit{e.g.,}  the spectrum efficiency.
For instance, consider a wireless system with $N$ transmitter and receiver pairs sharing the same bandwidth. Let $h_{ii} \in \mathbb{C}$ denote the  $i$-th channel, $h_{ij} \in \mathbb{C}$ denote the interference channel from the $i$-th transmitter to the $j$-th receiver, and $\sigma_i^2$ denote the noise power at the $i$-th receiver.
The optimal power $\{p_i, i = 1, ..., N\}$ for the maximum sum-rate is formulated as
\begin{equation}
    \max_{\{p_i, i = 1, ..., N\}} \sum_{i=1}^{N} \log \left(1 + \frac{|h_{ii}|^2 p_i}{\sum_{i \neq j} |h_{ij}|^2 p_j + \sigma_i^2}\right).
\end{equation}
This problem can be addressed via the decentralized optimization framework with
\begin{equation*}
f_i = \log \left(1 + \frac{|h_{ii}|^2p_i}{\sum_{i \neq j} |h_{ij}|^2 p_j + \sigma_i^2}\right).     
\end{equation*}
Compared to centralized resource allocation schemes, which collect the local information at a server, conducting the resource allocation in a decentralized manner leads to a smaller communication overhead.

\subsubsection{Decentralized training of machine learning models} 
A variety of machine learning approaches can be formulated as an optimization problem over the training data, (\textit{e.g.,}  classification or regression).
When the data is stored across multiple devices, the training object can be formulated as the decentralized optimization problem, where $f_i$ is the training loss on the subset data on the $i$-th device. 
The decentralized SGD and its variants have been utilized for improving the scalability of machine learning models in both datacenters and decentralized networks of devices.
In addition, its orthogonal benefits compared to centralized training is on protecting users’ privacy.

% \subsubsection{Multi-agent System} Constraint resource allocations.

% The decentralized variants of SGD and other optimization algorithms have recently been considered in machine learning both for improved scalability in datacenters [30] as well as for decentralized networks of devices [110, 392, 379, 54, 243, 253, 153].
% They consider undirected network graphs, although the case of directed networks (encoding unidirectional channels which may arise in real-world scenarios such as social networks or data markets) has also been studied in [30, 200]. The work of [392, 54] introduces fully
% decentralized algorithms to collaboratively learn a personalized model for each client by smoothing model
% parameters across clients that have similar tasks (i.e., similar data distributions). Zantedeschi et al. [431], multi-task learning. A blockchain is a distributed ledger shared among disparate users, making possible digital transactions, including transactions of cryptocurrency, without a central authority.
% The applications on IoT is promissing as  Industrial Internet of Things (IIoT) systems [4]
% are pushing toward massively dense and fully decentralized networks that do not rely upon a central server, and where the terms of
% energy, computational power, bandwidth and channel uses, is much lower than the cost of a long-range server connection.

\subsection{Notation}
In the subsequent discussion, we use $N$ and $d$ to denote the number of device and the dimension of the training parameters, respectively.   
We use $\mathbf{x}_i^{(t)}$ to denote the parameters on device $i$ at time step $t$. We further define the average
\begin{equation}
    \bar{\mathbf{x}}_{t} := \frac{1}{N} \sum_{i=1}^{N} \mathbf{x}_i^{(t)}.
\end{equation}
We also use matrix notation where it is more convenient, \emph{i.e.}
\begin{equation}
    \mathbf{X}_t := [\mathbf{x}_1^{(t)},..., \mathbf{x}_N^{(t)}]^T \in \mathbb{R}^{N \times d}.
\end{equation}
In addition, we use $\mathbf{X}_l^{(t)}$ to denote the $l$-th column of $\mathbf{X}_t$.
Similarly, we use $\mathbf{G}_t := [\mathbf{g}_1^{(t)},..., \mathbf{g}_N^{(t)}]^T$ and $\nabla \mathbf{F}_t := [\nabla f_1(\mathbf{x}_1),..., \nabla f_N(\mathbf{x}_N)]^T$ to denote the stochastic gradients and gradients, respectively, and $\mathbf{G}_l^{(t)}$ and $\nabla \mathbf{F}_l^{(t)}$ to denote the $l$-th column of $\mathbf{G}_t$ and $\nabla \mathbf{F}_t$ respectively. 
We use $\mathbf{1}$ to denote the vector of ones and $\mathbf{J} := \frac{1}{N}\mathbf{11^T}$.
We use $\|\mathbf{y}\|$ to denote the Euclidean norm of a vector $y$.
For a matrix $\mathbf{W}$, we will use $w_{i,j}$ or $\mathbf{W}[i, j]$  to denote the $i, j$-th entry.

% Throughout this paper, we define the concatenation of all local variables, stochastic gradients, and their average to organize the algorithm more clearly.
% \begin{align}
%     &\mathbf{X}_t := [\mathbf{x}_1^{(t)},..., \mathbf{x}_N^{(t)}]^T \\
%     &\mathbf{X}_l^{(t)} :=  \mathbf{X}_t \mathbf{e}_l\\
%     &\mathbf{G}_t := [\mathbf{g}_1^{(t)},..., \mathbf{g}_N^{(t)}]^T \\
%     &\mathbf{G}_l^{(t)} := \mathbf{G}_t \mathbf{e}_l\\
%     & \nabla \mathbf{F}_t := [\nabla f_1(\mathbf{x}_1),..., \nabla f_N(\mathbf{x}_N)]^T \\
%     & \nabla \mathbf{F}_l^{(t)} := \nabla \mathbf{F}_t \mathbf{e}_l \\
%     &\\
%     & \mathbf{J} := \frac{1}{N}\mathbf{11^T} 
% \end{align}

\section{Soft-DSGD} \label{sec:Method}
\begin{figure*}[t]
    \centering
    \includegraphics[width=150mm]{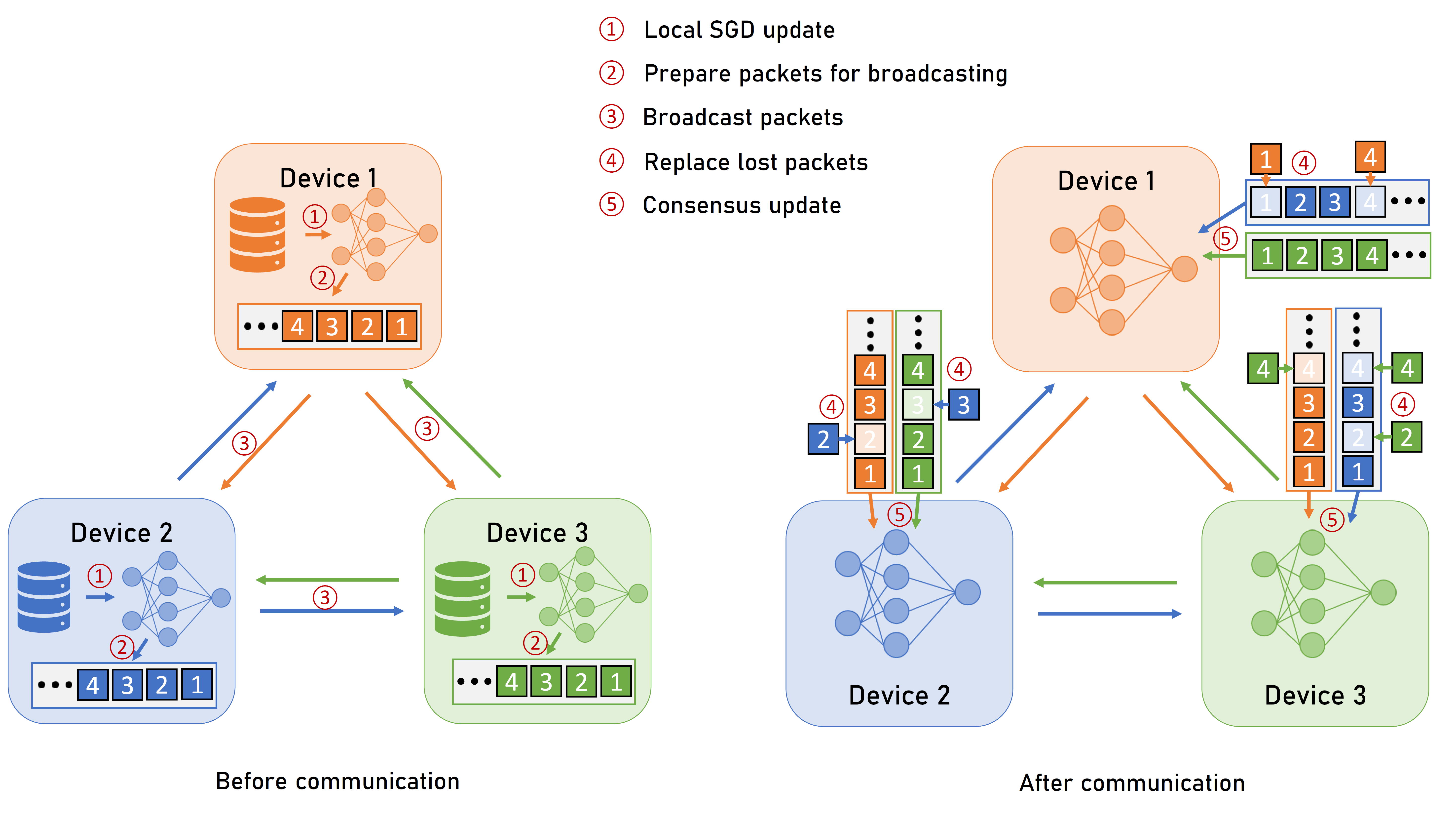}
    \caption{Pipeline of Soft-DSGD.} \label{fig:sys}
\end{figure*}

In this section, we introduce Soft-DSGD for training deep learning models with unreliable communication networks as shown in Fig \ref{fig:sys}.   

\subsection{Unreliable communications with UDP}

Most of the existing decentralized optimization methods require a reliable communication network among the devices while the real-world edge communication systems are prone to packet loss and transmission errors.
TCP and UDP are widely-used transportation layer protocols for transmitting data packets over a communication network.
TCP is a connection-oriented protocol and requires handshakes to set up an end-to-end communication connection, where messages can be transmitted in both directions.  TCP leverages mechanisms including ACK messages, retransmission, and timeouts to guarantee the transmission reliability.
The recipient sends ACK messages back to the sender once receiving the messages correctly. Otherwise, the transmitter will resend the packets to the recipient. 
When TCP is used in broadcast and multi-cast scenarios, reliability is guaranteed for each recipient. 
Hence, TCP has a considerable communication overhead, especially in unreliable communication networks, such as wireless systems. 
Compared to TCP, UDP is a lightweight message-based connectionless protocol with much less overhead and does not need to set up a dedicated connection for transmission. 
Communication is achieved by transmitting packets in one direction from the source to the destination without verifying whether the recipients have received the packets. Due to the unreliability nature of UDP, it is mainly used for delivery best effort traffic, such as multimedia streaming, where occasional packet loss can be tolerated.

In Soft-DSGD, an unreliable and connectionless UDP transmission protocol is adopted. 
Instead of modeling the communication network as a graph $\mathcal{G}$ where each device can only communicate with its neighbours along the edge as in TCP, we assume a communication network, where each device can send and receive messages from the rest of the network. 
However, the delivery is not guaranteed in UDP as the transmitted packets are exposed to unreliable communication conditions and no mechanism for reliability enhancement like packet retransmission is implemented.
We use a matrix $\mathbf{P} = [p_{i,j}] \in \mathbb{R}^{N\times N}$ to describe the level of link reliability in the communication network, where $p_{i,j}$ represents the probability of successful transmission from the $i$-th device to the $j$-th device and $p_{i,i}=0, \forall i \in \{1, ..., N\}$,. Parameter exchange among the devices consists of the following three steps as follows.
\begin{enumerate}
    \item \textbf{Dividing parameters into packets:} The number of parameters of machine learning models, especially deep neural networks, is usually too huge to be transmitted in a single packet. We assume that the parameters are randomly grouped into multiple packets and transmitted independently.
    \item \textbf{Broadcasting:}  The packets are broadcast to the rest of the network without targeted recipients. With UDP, it is unknown whether the packet can reach its destination with success.  
    \item \textbf{Stochastic receiving:} Due to the unreliability of communication links, packets may be dropped or contaminated randomly when received at any device. With the help of an error detection code, such as a checksum, the receiver can detect errors in the packet, in which case the packet is declared lost and  is discarded. Otherwise, the packet is successfully received. 
\end{enumerate}
In summary, after the $i$-th device broadcasts packets of its parameters $\mathbf{x}_i$ to the network, they are randomly received by nodes in the the rest of network. For instance, at the $j$-th device, the received data, $\mathbf{z}_{i\rightarrow j}$, may only include parts of $\mathbf{x}_i$. 
Therefore, it can be expressed as $\mathbf{z}_{i\rightarrow j} = \mathbf{m}_{i\rightarrow j} \odot \mathbf{x}_i$, where $\odot$ denotes element-wise multiplication and $\mathbf{m}_{i\rightarrow j}(k) = 1$ if the $k$-th parameter broadcast by the $i$-th device is successfully received at the $j$-th device and $\mathbf{m}_{i\rightarrow j}(k) = 0$ otherwise.

\subsection{Algorithm}
\begin{algorithm}[t]
\caption{Soft-DSGD Training Algorithm}
\begin{algorithmic}
\Require Initialize local models $\{\mathbf{x}_i\}$, learning rate $\gamma$, and mixing weight matrix $\mathbf{W}$.
\For {$t = 1$ to T} 
\For {$i=1$ to $N$}
\State Sample a mini-batch $\xi_i$ from local dataset $\mathcal{D}_i$.
\State Compute the local gradient with $\mathbf{g}_i^t = \nabla F_i(\mathbf{x}_i^{(t)}, \xi_i)$.
\State Local SGD update: $\mathbf{x}_i^{(t+\frac{1}{2})} = \mathbf{x}_i^{(t)} - \gamma \mathbf{g}_i^{(t)}$.
\State Broadcasting $\mathbf{x}_i^{(t+\frac{1}{2})}$ to the rest of network.
\State Receiving messages from other devices: $\mathbf{z}_{j \rightarrow i}^{(t+\frac{1}{2})} = \mathbf{m}_{j \rightarrow i }^{(t)} \odot \mathbf{x}_j^{(t+\frac{1}{2})}, j \neq i$.
\State Replace the missing values with stale data: $\hat{\mathbf{z}}_{j \rightarrow i}^{(t+\frac{1}{2})} = \mathbf{z}_{j \rightarrow i}^{(t+\frac{1}{2})} + (1 - \mathbf{m}_{j \rightarrow i }^{(t)}) \odot  \mathbf{x}_i^{(t+\frac{1}{2})}$.
\State Consensus update: $\mathbf{x}_i^{(t+1)} =w_{i, i} \mathbf{x}_i^{(t+\frac{1}{2})} + \sum_{j=1, j\neq i}^N w_{i, j} \hat{\mathbf{z}}_{j \rightarrow i}^{(t+\frac{1}{2})}$. 
\EndFor
\EndFor
\end{algorithmic} 
\end{algorithm}

Soft-DSGD, illustrated in Algorithm 1, is designed to address unreliability in UDP transmissions.
We adopt the vanilla decentralized training framework \cite{lian2017can}, where each device maintains its own local parameters and conducts a local SGD update as well as a consensus update in each training iteration.
The key challenges are how to conduct a consensus update with only partially received messages from  other devices and how to optimize  mixing weight matrix $\mathbf{W}$ according to link reliability matrix $\mathbf{P}$.

\textbf{Filling lost packets with local parameters.} 
%The received parameters $\mathbf{z}_{i\rightarrow j}$ at device $j$ has missed values due to the transmission failures. 
To deal with transmission failures in UDP, we replace the lost packets with local parameters at each device. In particular, the $i$-th device will fill the missing parameters in $\mathbf{z}_{j \rightarrow i}$ with those in $\mathbf{x}_i$ and the filled message $\mathbf{\hat{z}}_{j \rightarrow i}$ can expressed as 
\begin{equation}
    \hat{\mathbf{z}}_{j \rightarrow i}^{(t+\frac{1}{2})} = \mathbf{z}_{j \rightarrow i}^{(t+\frac{1}{2})} + (1 - \mathbf{m}_{j \rightarrow i}^{(t)}) \odot  \mathbf{x}_i^{(t+\frac{1}{2})}.
\end{equation}
With the filled parameters, the consensus update can be expressed as
\begin{align}
    \mathbf{x}_i^{(t+1)} &=w_{i,i} \mathbf{x}_i^{(t+\frac{1}{2})} + \sum_{j=1, j\neq i}^N w_{i,j} \hat{\mathbf{z}}_{j \rightarrow i}^{(t+\frac{1}{2})} \\
    & = \mathbf{x}_i^{(t+\frac{1}{2})} + \sum_{j=1, j\neq i}^N w_{i,j} \mathbf{m}_{j \rightarrow i }^{(t)} \odot( \mathbf{x}_j^{(t+\frac{1}{2})} - \mathbf{x}_i^{(t+\frac{1}{2})}).
\end{align}

Note that since we use the local parameters to replace the lost values, there is no additional memory required to store the historically received data.

Due to the randomness of the communication network, the consensus update step becomes stochastic. The following lemma, proved in Appendix, introduces two important matrices $\overline{\mathbf{W}}$ and $\overline{\mathbf{W}^2}$, which characterize the first and second moments of the update parameters and also play an important role in the analysis of the convergence rate.

\textbf{Lemma 1.} \textit{With the updating rule, the expectations of $\mathbf{X}_{t+1}$ and $(\mathbf{X}_l^{(t+1)})^T(\mathbf{X}_l^{(t+1)})$ can be expressed as} 
\begin{align}
    \mathbb{E}\{\mathbf{X}_{t+1}\} &= \overline{\mathbf{W}}(\mathbf{X}_t - \gamma \mathbf{G}_t),
\end{align}
and
\begin{align}
  \mathbb{E}\{ (\mathbf{X}_l^{(t+1)})^T(\mathbf{X}_l^{(t+1)}) \}&= (\mathbf{X}_l^{(t)} -  \gamma \mathbf{G}_l^{(t)})^T \overline{\mathbf{W^2}} (\mathbf{X}_l^{(t)} -  \gamma \mathbf{G}_l^{(t)}),
\end{align}
\textit{where $\overline{\mathbf{W}}$ and $\overline{\mathbf{W}^2}$ are defined as:}
% ============= Single Column ==========
% \begin{align}
%     \overline{\mathbf{W}}[i, j] &:= \begin{cases}
%      w_{i,j} p_{i,j},  \quad \quad \quad \quad \quad \quad  & i \neq j\\
%      1 - \sum_{l = 1, l\neq i}^N w_{i,l} p_{i,l}, \quad & i = j\\
%     \end{cases} \\
%     \overline{\mathbf{W}^2} [i, j] &:= \begin{cases}
%      \sum_{l = 1}^N w_{i,l} p_{i,l} w_{j,l} p_{j,l} 
%      + p_{i,j} w_{i,j}\left(2 - \sum_{l=1}^N \left( p_{i, l} w_{i, l} + p_{j, l} w_{j, l}\right)\right),  \quad & i \neq j\\
%      1 - 2\sum_{l=1}^N p_{i, l}\left(w_{i,l}- w_{i, l}^2\right)
%      + \sum_{l=1}^N \sum_{m=1, m\neq l}^N p_{i, l} w_{i, l} p_{i, m} w_{i, m}, \quad & i = j\\
%     \end{cases}
% \end{align}

% ============= Double Column ==========

\begin{align}
    \overline{\mathbf{W}}[i, j] &:= \begin{cases}
     w_{i,j} p_{i,j},  \quad \quad \quad \quad \quad \quad  & i \neq j\\
     1 - \sum_{l = 1, l\neq i}^N w_{i,l} p_{i,l}, \quad & i = j\\
    \end{cases} \\
    \overline{\mathbf{W}^2} [i, j] &:= \begin{cases}
     \sum_{l = 1}^N w_{i,l} p_{i,l} w_{j,l} p_{j,l} 
     + 2 p_{i,j} w_{i,j} \\
     -  p_{i,j} w_{i,j} \sum_{l=1}^N \left( p_{i, l} w_{i, l} + p_{j, l} w_{j, l}\right),  \quad & i \neq j\\
     1 - 2\sum_{l=1}^N p_{i, l}\left(w_{i,l}- w_{i, l}^2\right)\\
     + \sum_{l=1}^N \sum_{m=1, m\neq l}^N p_{i, l} w_{i, l} p_{i, m} w_{i, m}, \quad & i = j\\
    \end{cases}
\end{align}

\textbf{Remark 1.} Lemma 1 illustrates the first- and second-order of statistics of the soft-DSGD updating. In expectation, the consensus updates with an unreliable communication network are equivalent to reliable consensus updates with $\overline{\mathbf{W}}$ as the weight matrix, which is also a doubly stochastic matrix.

\textbf{Optimizing mixing matrix.}
%In the vanilla decentralized training framework, the mixing matrix $\mathbf{W}$ is set according to the topology of the communication network \cite{h}. 
%With unreliable communication, the convergence are determined by both the mixing matrix $\mathbf{W}$ and the transmission matrix $P$.
In vanilla decentralized SGD, the convergence rate of training largely depends on the mixing matrix $\mathbf{W}$. From Lemma 1, the average mixing weight $\overline{\mathbf{W}}$ depends not only on $\mathbf{W}$  but also on the link reliability matrix $\mathbf{P}$. In this paper, two approaches are exploited to select the mixing matrix $\mathbf{W}$ , depending on the availability of the matrix $\mathbf{P}$.  
\begin{enumerate}
    \item If link reliability matrix $\mathbf{P}$ is unknown, each link will be treated equally and uniform mixing weights will be adopted, \emph{i.e.}, $\mathbf{W} = \mathbf{J} = \frac{1}{N} \mathbf{11}^T$.
    \item If link reliability matrix $\mathbf{P}$ is available, (\emph{e.g.,} maintained at a coordinator), we can optimize $\mathbf{W}$ for faster convergence. As will be shown in the next section, the convergence of Soft-DSGD depends on the largest eigenvalue of matrix $\overline{\mathbf{W}^2} - \mathbf{J}$. Therefore, we optimize $\mathbf{W}$ to minimize the largest eigenvalue of this matrix:
    \begin{align}
    \min_{\mathbf{W}} \quad &\lambda_{\max}(\overline{\mathbf{W}^2} - \mathbf{J}) \\
    \textrm{s.t.} \quad &  0 \leq w_{i, j} \leq 1,  \mathbf{W}^T = \mathbf{W}, \mathbf{W} \mathbf{1}  = \mathbf{1} 
    \end{align}
    In fact, this optimization problem is convex, as shown in the Appendix.  Therefore, it can be solved efficiently.
\end{enumerate}

\section{Convergence Analysis} \label{sec:Theory}
In this section, we analyze convergence of Soft-DSGD and prove that even in unreliable communication networks, Soft-DSGD  achieves the same asymptotic convergence rate as the vanilla decentralized SGD with perfect communications.

\subsection{Assumptions}

\textbf{Assumptions on functions.} The functions $f$ and $f_i$ have following properties.
\begin{itemize}
    \item \textbf{($L$-smoothness)}. \textit{Each local function $f_i(\cdot)$ is smooth and with $L$-Lipschitzian gradients, \emph{i.e.}, there exists a constant $L > 0$, such that  $\forall \mathbf{x}, \mathbf{y} \in \mathbb{R}^d$, }
   \begin{equation}
    \| \nabla f_i(\mathbf{x}) - \nabla f_i(\mathbf{y}) \| \leq L \|\mathbf{x} - \mathbf{y}\|.
    \end{equation}
    \item \textbf{(Bounded variance).} \textit{We assume that there exists constants $\sigma > 0$ and $\zeta > 0$, such that $\forall \mathbf{x} \in \mathbb{R}^d$,}
\begin{align}
    \mathbb{E} \{ \| \nabla F_i(\mathbf{x}, \xi) - \nabla f_i(\mathbf{x})\| \} \leq \sigma^2,\\
    \frac{1}{N} \sum_{i=1}^{N} \| \nabla f_i(\mathbf{x}) - \nabla f(\mathbf{x}) \| \leq \zeta^2.
\end{align}
Hence, $\sigma^2$ bounds the variance of stochastic gradients at each device and $\zeta^2$ bounds the discrepancy of data distributions at different devices. 
\item \textbf{(Unbiased stochastic gradients)}. Stochastic gradients obtained at each device are unbiased estimates of the real gradients of the local objectives:
\begin{equation}
    \mathbb{E} \{\mathbf{g}_i\} = \nabla f_i(\mathbf{x}_i).
\end{equation}

\end{itemize}

These assumptions on functions are widely used in the non-convex decentralized optimization literature \cite{lian2017can,tang2018communication,koloskova2019decentralized}, and are valid in most of applications.

\textbf{Assumptions on communication networks.}  Besides the above assumptions on the functions, we make additional assumptions on the unreliable communication network.
\begin{itemize}
    \item \textbf{(Symmetric matrix)}. \textit{The probability for successful transmission from the $i$-th device to the $j$-th device is the same as the probability from the $j$-th device to the $i$-th device, \emph{i.e.,} $\mathbf{P}^T = \mathbf{P}$.}
    \item \textbf{(Independent and stable links)}. \textit{The packet transmission on different links are independent and the link reliability matrix $\mathbf{P}$ remains fixed during training.}
\end{itemize}

These assumptions on communication networks are reasonable and easy to be satisfied in practice. 
Due to channel reciprocity, the link reliability is the same for transmissions in two directions on the same communication link. 
In addition, the assumption on independence of the links is valid as long as the distance between devices are much larger than the wavelength of the signal and the link reliability remains stable if the devices are static during the training.
Note that Soft-DSGD can be directly applied in the dynamic environment, where the link reliability matrix changes with time. We concentrate on the static reliability matrix in order to make the analysis easy to understand.

\subsection{Soft-Consensus Algorithm}
To analyze the convergence of Soft-DSGD, we first investigate the consensus algorithm with an unreliable communication network.  
Suppose the devices are initialized with $\{ \mathbf{x}_i^{(0)} \in \mathbb{R}^d, i=1,...,N\}$, and only the consensus update steps are conducted in each iteration, \emph{i.e.},
\begin{equation}
    \mathbf{x}_i^{(t+1)} = \mathbf{x}_i^{(t)} + \sum_{j=1, j\neq i}^N w_{i,j} \mathbf{m}_{i \rightarrow j }^{(t)} \odot( \mathbf{x}_i^{(t)} - \mathbf{x}_j^{(t)} )
\end{equation}

Then, we have the following lemma, proved in Appendix, to capture the resistance of the random communication network.

\textbf{Lemma 2.} \textit{Let $\overline{\mathbf{x}}_t$ denote the average of $\mathbf{X}_t$.  Following the consensus updating rule,  we have}
\begin{align}
    \mathbb{E} \{\overline{\mathbf{x}}_{t+1}\} &= \overline{\mathbf{x}}_t, \\ 
    \mathbb{E}\{ \| \overline{\mathbf{x}}_{t+1}  - \overline{\mathbf{x}}_t \|^2\} &\leq  \frac{\kappa}{N^2}\sum_{i=1}^N \|\mathbf{x}_i^{(t)} - \overline{\mathbf{x}}_t\|^2,
\end{align}
\textit{where the expectation is taken over the randomness of the communication network and $\kappa = 2 \max_i \sum_{j=1}^N w^2_{i,j} p_{i,j}(1 - p_{i,j}) $. }

\textbf{Remark 2.} Lemma 2 shows how the average of $\mathbf{X}_t$ behaves with the consensus updates. In particular,  the average is preserved in expectation for each step and the expected deviation is bounded by the variance of $\mathbf{X}_t$. $\kappa$ captures the resistance of the random communication network. If the communication links are deterministic, \emph{i.e.,} $p_{i,j} = 0$ or $1$, $\kappa$ becomes zero. In this case, the average of $\mathbf{X}_t$ will be preserved for each step.

The convergence of the decentralized SGD depends on the convergence rate of the consensus. The convergence rate of consensus with unreliable communications is shown in the following lemma, proved in Appendix.

\textbf{Lemma 3.} \textit{Following the consensus updating rule,  the expectation of $\frac{1}{N} \sum_{i=1}^N \|\mathbf{x}_i^{(t)} - \overline{\mathbf{x}}_t \|^2$ converges to zero at an exponential rate. In particular, we have}
\begin{equation}
    \mathbb{E} \{ \frac{1}{N} \sum_{i=1}^N \|\mathbf{x}_i^{(t)} - \overline{\mathbf{x}}_t \|^2 \} \leq \rho^t \sum_{i=1}^N \| \mathbf{x}_i^{(0)} - \overline{\mathbf{x}}_0 \|^2,
\end{equation}
\textit{where $\rho$ is the largest eigenvalue of matrix $\overline{\mathbf{W}^2} - \mathbf{J}$}.

\textbf{Remark 3.}  If the communication network is reliable, the consensus enjoys an exponential convergence rate. Lemma 3 shows that the consensus can also achieve an exponential convergence rate with unreliable communications under the proposed consensus update policy.

\subsection{Convergence}
Based on the above assumptions and lemmas, the convergence rate for the proposed decentralized training algorithm with an unreliable communication network can be demonstrated in the following theorem, proved in the appendix.

\textbf{Convergence of Soft-DSGD Theorem}. \textit{Suppose that all local models are initialized with $\mathbf{x}_0 \in \mathbb{R}^d$. Under Assumptions 1-5, if the learning rate satisfies $\gamma L \leq \min\{1, (\sqrt{\rho^{-1}} -1)/4 \}$, then after $T$ iterations, we have}
% ======= Single Column =======
% \begin{align}
%      \frac{1}{T}\sum_{t=1}^T \| \nabla f(\bar{\mathbf{x}}_t)\|^2 
%      &\leq \left(\frac{\mathbb{E}[f(\bar{\mathbf{x}}_{T})] - \mathbb{E}[f(\bar{\mathbf{x}}_0)]}{\gamma T} + \frac{\gamma L}{N}\sigma^2 + \frac{2 \gamma L\kappa}{N}\sigma^2 + \frac{6L\kappa\gamma \zeta^2}{N}\right) \frac{1-D}{1-2D} \\
%      &+ \left(L^2 + \frac{2 L\kappa}{\gamma N} + \frac{2(3N+1)L^3\gamma\kappa}{N}\right)\left(\frac{2\gamma^2\sigma^2\rho}{1-\rho}  +\frac{6\gamma^2\zeta^2\rho}{(1-\sqrt{\rho})^2}\right) \frac{1}{1-2D},
% \end{align}
% ======= Double Column =======
\begin{align}
      \frac{1}{T}\sum_{t=1}^T \| \nabla f(\bar{\mathbf{x}}_t)\|^2 & \leq  (\frac{\mathbb{E}[f(\bar{\mathbf{x}}_0)] - \mathbb{E}[f(\bar{\mathbf{x}}_{T})]}{\gamma T} \nonumber \\
     & + \frac{\gamma L}{N}\sigma^2 + \frac{2 \gamma L\kappa}{N}\sigma^2 + \frac{6L\kappa\gamma \zeta^2}{N} ) \frac{1-D}{1-2D} \nonumber \\
     & + \left(L^2 + \frac{2 L\kappa}{\gamma N} + \frac{2(3N+1)L^3\gamma\kappa}{N}\right) \nonumber\\
     & \left(\frac{2\gamma^2\sigma^2\rho}{1-\rho}  +\frac{6\gamma^2\zeta^2\rho}{(1-\sqrt{\rho})^2}\right) \frac{1}{1-2D},
\end{align}
\textit{where $D =\frac{6\gamma^2 L^2 \rho}{(1-\sqrt{\rho})^2}$, $\kappa = 2 \max_i \sum_{j=1}^N p_{i,j}\left(1 - p_{i,j}\right) w_{i,j}$, and $\rho$ is the largest eigenvalue of the matrix $\overline{\mathbf{W}^2} - \mathbf{J}$.}

The resistance of the unreliable communication network is reflected in terms containing $\kappa$. If all the communication links are deterministic with $p_{i,j}=0$ or $1$, then $\kappa = 0$, and the results will be consistent with the convergence bound for vanilla decentralized SGD.
In addition, the convergence bound depends on $\rho$ to a large degree, which justifies our mixing weight optimization method.

Furthermore, if the learning rate is configured properly, it can achieve a linear speedup in terms of the number of devices, matching the same rate as vanilla decentralized SGD, as indicated in the following corollary, proved in Appendix.

\textbf{Corollary}. Under the same conditions as the above theorem, if the learning rate $\gamma$ is set as $\gamma = \sqrt{\frac{N}{T}}$, after total $T$ iterations, we have
\begin{equation}
     \frac{1}{T}\sum_{t=1}^T \| \nabla f(\bar{\mathbf{x}}_t)\|^2 = \mathcal{O}(\frac{1}{\sqrt{NT}}) + \mathcal{O}(\frac{N}{T}),
\end{equation}
where all other constants are subsumed in $\mathcal{O}$.

\textbf{Consistency with vanilla decentralized SGD}. Recall that vanilla decentralized SGD converges at the asymptotic rate of $\mathcal{O}(\frac{1}{\sqrt{NT}}) + \mathcal{O}(\frac{N}{T})$ \cite{lian2017can}. Hence, the decentralized SGD with unreliable communications can achieve the same asymptotic convergence rate as vanilla decentralized SGD that assumes a reliable communication network. Therefore, the asymptotic convergence is not negatively affected by unreliability in the communication network.

\section{Experiments} \label{sec:Exp}
In this section we evaluate the performance of Soft-DSGD with unreliable communication networks.

% \begin{figure*}
%     \subfigure[]{\includegraphics[width=0.31\textwidth]{Figures/2a.png}} 
%     \subfigure[]{\includegraphics[width=0.31\textwidth]{Figures/2b.png}} 
%     \subfigure[]{\includegraphics[width=0.31\textwidth]{Figures/2c.png}}
%     \caption{(a) A communication graph generated with $p_\delta = 0.7$. (b) Training loss vs epochs. (c) Training loss vs communication rounds.}
% \end{figure*}

\subsection{Experimental setup}

\begin{figure}[t]
    \centering
    \includegraphics[width=0.49\textwidth]{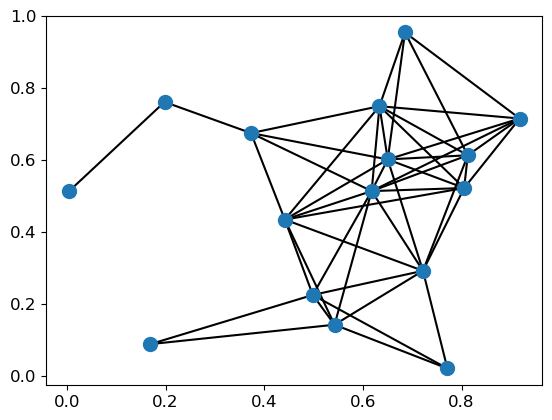}
    \caption{A communication graph generated with $p_\delta = 0.7$.} \label{fig:ComGraph}
\end{figure}

We conduct experiments on image classification tasks for evaluation. We train ResNet-20 models \cite{he2016deep} on the CIFAR-10 dataset \cite{krizhevsky2009learning}, which contains $50,000$ images for training and $10,000$ images for testing. We set the weight decay to $0.0001$ and mini-batch size to $32$ per device. The initial learning rate is $0.1$ and decays by a factor of $10$ after $60$ epochs. The momentum of $0.9$ is used.

%The communication network is set to be unreliable. Given $N$ device, there are $\frac{N(N-1)}{2}$ links in all and the packet drop rates of the links are set to follow a uniform distribution in interval $[0, 1]$.

Geometric random graphs are generated to represent a communication network. The network consists of $16$ devices, which are randomly located in a unit square as shown in Fig \ref{fig:ComGraph}. 
The probability of successful transmission for each link is defined in a way that decays with the distance between the devices, \emph{i.e.,}     $p_{i,j} =  p_{j, i} = k^{(\frac{d_{i,j}}{r})^2}$,
where $d_{i,j}$ represents the distance of the $i$-th device to the $j$-th device and we set $k = 0.7$ and $r = 0.4$. For example, the transmission success probability is $0.7$ when the distance between two devices is $0.4$.

We implement Soft-DSGD with PyTorch and train models with an Nvidia 1080Ti GPU.
%The training process takes about $12$ hours for $100$ epochs, where the computation complexity mostly lies in the simulation of the unreliable communication network. 
To simulate the random communication network, a random mask $\mathbf{m}_{i \rightarrow j}$ is generated before each communication to determine which part of the data be obtained by the receiver.
Since the packet size ($\sim 10^2$) is usually much smaller compared to the number of parameters ($\sim10^6$) and the random partition of the parameters into packets can be different for each device and time step as long as the random seeds are available for the receivers, the communications for each dimension of parameters can be approximated to be independent.  We simulate $\mathbf{m}_{i \rightarrow j}$ by sampling from $i.i.d.$ Bernoulli distribution regardless of the packet size.

\begin{figure}[t]
    \subfigure[]{\includegraphics[width=0.49\textwidth]{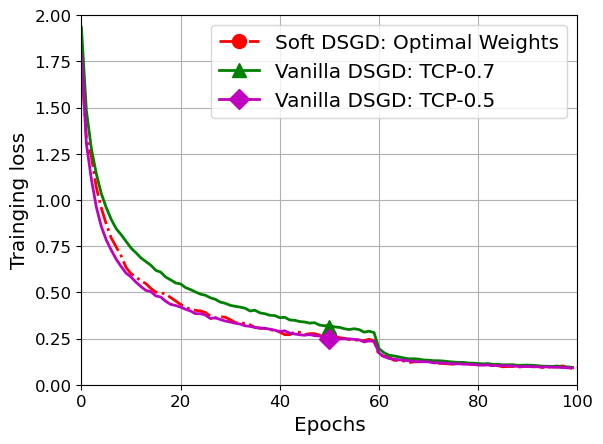}}
    \subfigure[]{\includegraphics[width=0.49\textwidth]{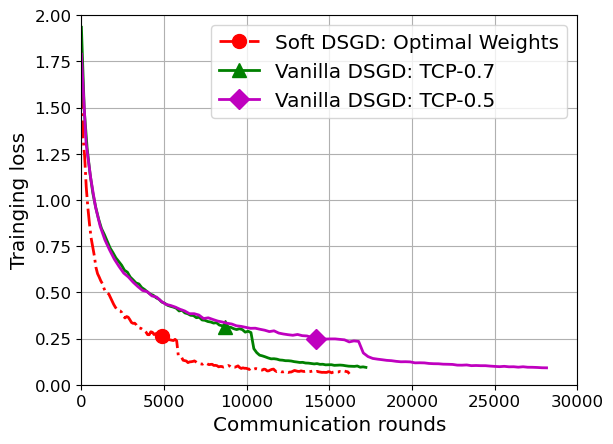}}
    \caption{(a) Training loss vs epochs. (b) Training loss vs communication rounds.} 
    \label{fig:res_comm}
\end{figure}

The proposed approach is compared with vanilla decentralized SGD using TCP, where a communication graph $\mathcal{G}(V, E)$ is constructed first and the devices only exchange information with their neighbours.
The neighbours of a device are determined by a probability threshold $p_{\delta}$. 
Only links with a success probability larger than $p_{\delta}$ are kept while other links are discarded.
%Fig. \ref{fig:ComGraph} shows positions of the devices used in the experiments and a TCP communication graph with $p_{\delta} = 0.7$.
After the communication graph $\mathcal{G}(V, E)$ is constructed, the Metropolis-Hastings mixing weights are employed \cite{boyd2006randomized}, \emph{i.e.,}
% ===== Single Column =======
% \begin{equation*}
%     w_{i,j} = \begin{cases}
%     1/(\max\{ \text{deg}(i), \text{deg}(j)\} + 1) \quad &\text{$i \neq j$, $i$ and $j$ are neighbours in $\mathcal{G}$},\\
%     0 \quad &\text{$i \neq j$, $i$ and $j$ are not neighbours in $\mathcal{G}$},  \\
%     1 - \sum_{l = 1, l \neq i}^N 1/(\max\{ \text{deg}(i), \text{deg}(l)\} + 1) \quad & i = j.\\
%     \end{cases}
% \end{equation*}
% ===== Double Column =======
\begin{equation*}
    w_{i,j} = \begin{dcases*}
    0, \quad \quad \quad \quad \quad \quad \quad \quad \quad \quad \quad \, \, \, \, \text{$i \neq j$},  \text{$(i, j) \notin E $},  \\
    \frac{1}{(\max\{ \text{deg}(i), \text{deg}(l)\} + 1)},  \quad \text{$i \neq j$}, \text{$(i, j) \in E$},\\
    1 - \sum_{l = 1, l \neq i}^N \frac{1}{(\max\{ \text{deg}(i), \text{deg}(l)\} + 1)},  \, \,\, \,\text{$i = j$}.\\
    \end{dcases*}
\end{equation*}
where deg$(i)$ denotes the number of neighbours of the $i$-th device. %\textcolor{red}{where deg()...}

With TCP as the communication protocol, the receiver will send the ACK to the transmitter once it  successfully receives the packet. Otherwise, the transmitter will resend the last packet. If there are multiple neighbors, the transmitter needs to collect the ACK messages from all its neighbours to ensure reliability.

\begin{figure}[t]
    \centering
    \includegraphics[width=0.49\textwidth]{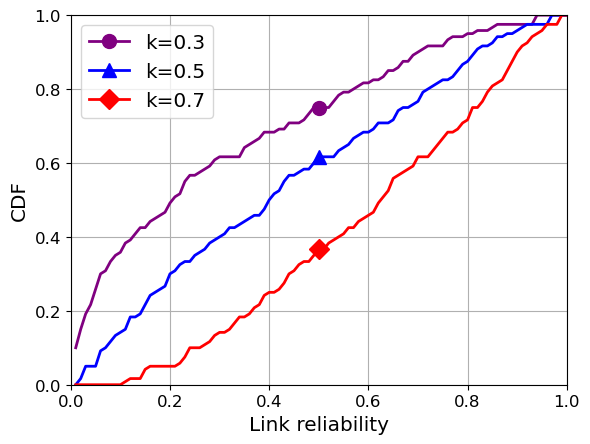}
    \caption{CDF of link reliability.}
    \label{fig:CDF}
\end{figure}

\subsection{Empirical results}

\paragraph{Effectiveness of Soft-DSGD} Fig. \ref{fig:res_comm} (a) compares the iteration-wise training convergence of Soft-DSGD and vanilla decentralized SGD with $p_\delta = 0.5$ and $p_\delta = 0.7$. A smaller $p_\delta$ leads to a denser communication network and requires fewer epochs to converge.  From this figure, the training of Soft-DSGD has a similar convergence with vanilla decentralized SGD with threshold $0.5$ and a much faster convergence than vanilla decentralized SGD with threshold $0.7$. 
This is because Soft-DSGD leverages information from all unreliable communication links and updates with the partially received messages while vanilla decentralized SGD only updates with messages from a limited number of neighbours.

Fig. \ref{fig:res_comm} (b) shows the convergence with respect to the communication round. For Soft-DSGD, the communication round is the same as the number of iterations, while for TCP, the packets need to be retransmitted in case of packet loss or transmission errors. Therefore the required communication round is greater than the Soft-DSGD. In addition, with a smaller $p_\delta$, the average number of neighbours increase while the average quality of communication links decrease, which results in worse performance in terms of required communication.

\begin{figure}[t]
    \subfigure[]{\includegraphics[width=0.49\textwidth]{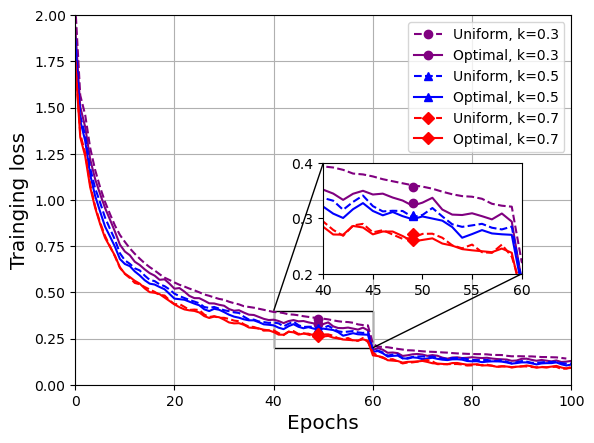}} 
    \subfigure[]{\includegraphics[width=0.49\textwidth]{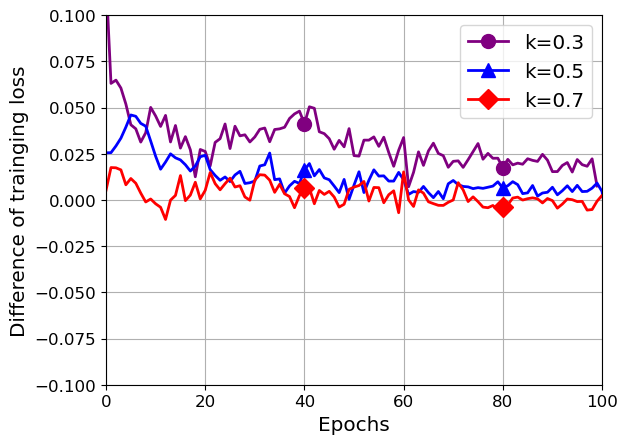}}
    \caption{ (a) Training loss vs epochs. (b) Training loss difference of uniform and optimal weights vs epochs.}
    \label{fig:res_compare}
\end{figure}

\paragraph{Effects of the link reliability matrix} We further investigate the performance of Soft-DSGD with different link reliability matrices.  We keep the positions of devices fixed and change $k$ to generate different link reliability matrices. The cumulative distribution function (CDF) of the link reliability is shown in Fig. \ref{fig:CDF} with different $k$. Fig. \ref{fig:res_compare} (a) shows the training loss of Soft-DSGD with different link reliability matrices, where the training slows down with the degradation of the communications links. We do not include vanilla SGD here because the communication graph are not connected for TCP when $k = 0.3$, but Soft-DSGD still works well.

To compare two weights setting approaches, Fig. \ref{fig:res_compare} (b) illustrates the difference of training losses between uniform and optimal mixing weights. When $k = 0.7$ and the average reliability of links is high, the performance gap between the uniform and optimal weights is quite small. But as the links quality degrades, the performance gap between two approaches increases. This is because when there are many links with little probability of transmission success, uniform weights still assign equal weights to these links, which impedes the convergence.

% \begin{table}[h]
%     \centering
%     \begin{tabular}{|c|c|c|c|}
%     \hline
%       \multicolumn{2}{|c|}{k=0.3}  &  \multicolumn{2}{|c|}{k=0.7} \\
%      \hline
%      Uniform weights & Optimal weights & Uniform weights & Optimal weights \\
%      \hline
%      90.00 \% & 89.97\% & 90.04\% & 90.02\% \\
%      \hline
%     \end{tabular}
%     \caption{Test accuracy of Soft-DSGD.}
%     \label{tab:my_label}
% \end{table}

% \paragraph{Test accuracy} The test performance of Soft-DSGD is shown in Figure \ref{fig:app}(b) and Table 1. The test accuracy of Soft-DSGD on a more reliable network ($k=0.7$) is slightly better than Soft-DSGD on a degraded network ($k=0.3$).

\section{Conclusion and Future Work} \label{sec:Conclusion}
In this paper, we have designed a robust decentralized training framework, called Soft-DSGD, to deal with unreliable communication links. 
Instead of using connection-orientated protocols, such as TCP, to ensure the reliability but with considerable overhead, Soft-DSGD uses lightweight and unreliable communication protocols, such as UDP, with low communication overhead, but is robust to communication failures.
We prove that Soft-DSGD, even with unreliable communications,  converges at the same asymptotic rate as the vanilla decentralize SGD over a reliable communication network.
From the numerical experiments, Soft-DSGD can leverage information collected from all unreliable communication links to accelerate  convergence.

\appendix

\subsection{Proof of Lemma 1}
\begin{proof}
$\mathbf{X}_l^{(t+1)} $, the $l$-th column of $\mathbf{X}_{t+1}$, can be expressed as $\mathbf{X}_l^{(t+1)}  = \widetilde{\mathbf{W}}_l^{(t)} \left( \mathbf{X}_l^{(t)}  - \gamma \mathbf{G}_l^{(t)}\right)$, where $\widetilde{\mathbf{W}}_l^{(t)}  \in \mathbb{R}^{N \times N}$ is the mixing matrix for $l$-th dimension.
$\widetilde{\mathbf{W}}_l^{(t)} $ is obtained by $\widetilde{\mathbf{W}}_l^{(t)}  = \mathbf{W} \odot \mathbf{A}_l^{(t)} + \mathbf{I} - \text{Diag}(\mathbf{W}\mathbf{A}_l^{(t)})$, where $\mathbf{A}_l^{(t)} \in \mathbb{R}^{N \times N} $ denotes and sucessfulness of transmission. If the transmission of the $l$-th parameter from $i$ to $j$ is successful, then $\mathbf{A}_l^{(t)}[i, j] = 1$ and otherwise, $\mathbf{A}_l^{(t)}[i, j] = 0$.
Since the link reliability from $i$ to $j$ is $p_{i,j}$, the expectation of $\widetilde{\mathbf{W}}_l^{(t)}[i, j]$ is $p_{i,j}w_{i,j}$ and the expectation of $\widetilde{\mathbf{W}}_l^{(t)} [i, i] = 1 - \sum_{j=1}^N p_{i,j}w_{i,j}$.
Therefore, $\mathbb{E} \{\widetilde{\mathbf{W}}_l^{(t)} \} = \overline{\mathbf{W}}$ and
\begin{align*}
    \mathbb{E} \{\mathbf{X}_l^{(t+1)}\} &=  \mathbb{E} \{\widetilde{\mathbf{W}}_l^{(t)} \} \left( \mathbf{X}_l^{(t)} - \gamma \mathbf{G}_l^{(t)} \right)  \\
    &= \overline{\mathbf{W}}\left( \mathbf{X}_l^{(t)} - \gamma \mathbf{G}_l^{(t)} \right). 
\end{align*}
By combining all dimensions, we have 
\begin{equation*}
    \mathbb{E}\{\mathbf{X}_{t+1}\} = \overline{\mathbf{W}} \left( \mathbf{X}_{t} - \gamma \mathbf{G}_{t} \right). 
\end{equation*}
Similarly, we have 
\begin{align*}
    &\mathbb{E}\{ (\mathbf{X}_l^{(t+1)})^T(\mathbf{X}_l^{(t+1)}) \} \\
    & = (\mathbf{X}_l^{(t)} - \gamma\mathbf{G}_l^{(t)})^T \mathbb{E} \{ (\widetilde{\mathbf{W}}_l^{(t)})^T (\widetilde{\mathbf{W}}_l^{(t)}) \} (\mathbf{X}_l^{(t)} - \gamma \mathbf{G}_l^{(t)}).
\end{align*}
Given $\widetilde{\mathbf{W}}_l^{(t)}  = \mathbf{W} \odot \mathbf{A}_l^{(t)} + \mathbf{I} - \text{Diag}(\mathbf{W} \mathbf{A}_l^{(t)})$, we have
\begin{align*}
    & (\widetilde{\mathbf{W}}_l^{(t)})^T (\widetilde{\mathbf{W}}_l^{(t)}) \\
    & = (\mathbf{W} \odot \mathbf{A}_l^{(t)})^2 + \text{Diag}^2(\mathbf{W} \mathbf{A}_l^{(t)}) + \mathbf{I} + 2\mathbf{W} \odot \mathbf{A}_l^{(t)} - \text{Diag}(\mathbf{W} \mathbf{A}_l^{(t)}) \\
    &   - (\mathbf{W} \odot \mathbf{A}_l^{(t)})\text{Diag}(\mathbf{W} \mathbf{A}_l^{(t)})  - \text{Diag}(\mathbf{W}\mathbf{A}_l^{(t)}) (\mathbf{W} \odot \mathbf{A}_l^{(t)})
\end{align*}
Taking expectation, we can get
\begin{equation*}
    \overline{\mathbf{W}^2} = \mathbb{E} \{ (\widetilde{\mathbf{W}}_l^{(t)})^T (\widetilde{\mathbf{W}}_l^{(t)})\}. 
\end{equation*}
This concludes the proof of Lemma 1. 
\end{proof}
\subsection{Proof of Lemma 2}

\begin{proof}
(a). According to Lemma 1, if $\mathbf{G}_t$ is zero, then
\begin{equation*}
\mathbb{E}\{\mathbf{X}_{t+1}\} = \overline{\mathbf{W}} \mathbf{X}_{t}.
\end{equation*}
Since $\overline{\mathbf{W}}$ is a doubly stochastic matrix, $\mathbb{E} \{\overline{\mathbf{x}}_{t+1}\} = \frac{1}{N}\mathbf{1}^T \mathbb{E}\{\mathbf{X}_{t+1}\} =  \frac{1}{N} \mathbf{1}^T \overline{\mathbf{W}} \mathbf{X}_{t} =  \frac{1}{N} \mathbf{1}^T  \mathbf{X}_{t} = \overline{\mathbf{x}}_{t}$.
(b). We have
\begin{align*}
    & \| \frac{1}{N} \sum_{i=1}^N \mathbf{x}_i^{(t+1)} -\frac{1}{N} \sum_{i=1}^N \mathbf{x}_i^{(t)}) \|^2 \nonumber \\
    %& =   \| \frac{1}{N} \sum_{i=1}^N \left(\mathbf{x}_i^{(t)} + \sum_{j=1, j\neq i}^N  w_{i,j}\mathbf{m}_{i \rightarrow j }^{(t)} \odot (\mathbf{x}_j^{(t)} - \mathbf{x}_i^{(t)})\right) -\frac{1}{N} \sum_{i=1}^N \mathbf{x}_i^{(t)}\|^2 \\
    & = \| \frac{1}{N} \sum_{i=1}^N \sum_{j=1, j\neq i}^N  w_{i,j}\mathbf{m}_{i \rightarrow j }^{(t)} \odot (\mathbf{x}_j^{(t)} - \mathbf{x}_i^{(t)})) \|^2 \\
    & = \| \frac{1}{N} \sum_{i=1}^N \sum_{j=1, j\neq i}^N w_{i,j}\mathbf{m}_{i \rightarrow j }^{(t)} \odot ((\mathbf{x}_j^{(t)}-\bar{\mathbf{x}}_t) - (\mathbf{x}_i^{(t)}-\bar{\mathbf{x}}_t)) \|^2 \\
    & = \| \frac{1}{N} \sum_{i=1}^N (\mathbf{x}_i^{(t)}-\bar{\mathbf{x}}_t) \odot \left(\sum_{j=1, j\neq i}^N  w_{i,j} \left(\mathbf{m}_{i \rightarrow j }^{(t)}  - \mathbf{m}_{j \rightarrow i }^{(t)}  \right)\right) \|^2. 
\end{align*}
Each item in $\mathbf{m}_{i \rightarrow j }^{(t)} - \mathbf{m}_{j \rightarrow i }^{(t)}$ is a variable with mean zero and variance of $2p_{i,j}(1-p_{i,j})$. Therefore, we have
\begin{align*}
    &\mathbb{E}\{\| \overline{\mathbf{x}}_{t+1}  - \overline{\mathbf{x}}_t \|^2\} \\
    &= \frac{1}{N^2} \sum_{i=1}^N \left( \|\mathbf{x}_i^{(t)}-\bar{\mathbf{x}}_t\|^2\sum_{j=1, j \neq i}^N 2 w_{i,j}^2 p_{i,j}(1-p_{i,j}) \right)\\
    & \leq \frac{\kappa}{N^2} \sum_{i=1}^N \|\mathbf{x}_i^{(t)}-\bar{\mathbf{x}}_t\|^2.
\end{align*}
This concludes the proof of Lemma 2.
\end{proof}
\subsection{Proof of Lemma 3}
\begin{proof}
We first consider the $l$-th dimension. 
Let $\bm{\beta}_l(t) = (\mathbf{I} - \mathbf{J}) \mathbf{X}_l^{(t)}$. 
Then we have
\begin{equation*}
   \bm{\beta}_l(t)  = (\widetilde{\mathbf{W}}_l^{(t)} - \mathbf{J}) \bm{\beta}_l(t-1). 
\end{equation*}
Now, taking the expected norm of $\bm{\beta}_l(t)$ given $\bm{\beta}_l(t-1)$, we have
\begin{align*}
    &\mathbb{E} \{ \| \bm{\beta}_l(t) \|^2 | \bm{\beta}_l(t-1)\}\\
    &=  \bm{\beta}_l(t-1)^T \mathbb{E}\{(\widetilde{\mathbf{W}} _l^{(t)}- \mathbf{J})^T (\widetilde{\mathbf{W}}_l^{(t)}  - \mathbf{J}) \} \bm{\beta}_l(t-1)  \\
    &= \bm{\beta}_l(t-1)^T \left( \overline{\mathbf{W}^2} - \mathbf{J}\right) \bm{\beta}_l(t-1)\\
    & \leq \rho \|\bm{\beta}_l(t-1)\|^2.
\end{align*}

Combine all dimensions together and let $\bm{\beta}(t) = (\mathbf{I} - \mathbf{J})\mathbf{X}_t$,
hence $\|\bm{\beta}(t)\|_F^2 = \sum_{i=1}^N \|\mathbf{x}_i^{(t)} - \overline{\mathbf{x}}_t \|^2$. Then we have
\begin{equation*}
    \mathbb{E} \{ \| \bm{\beta}(t) \|^2 | \bm{\beta}(t-1)\} \leq \rho \|\bm{\beta}(t-1)\|^2.
\end{equation*}
Repeat the above procedure over $t$, then we have 
\begin{equation*}
     \mathbb{E} \{ \|\bm{\beta}(t)\|^2 \}\leq \rho^t  \|\bm{\beta}(0)\|^2.
\end{equation*}
\end{proof}

\subsection{Proof of Theorem}

\begin{proof} Due to the Lipschitz smoothness of $f$, we have
\begin{equation*}
    f(\bar{\mathbf{x}}_{t+1}) - f(\bar{\mathbf{x}}_t) \leq \langle \nabla f(\bar{\mathbf{x}}_t), \bar{\mathbf{x}}_{t+1} - \bar{\mathbf{x}}_t\rangle + \frac{L}{2} \| \bar{\mathbf{x}}_{t+1} - \bar{\mathbf{x}}_t\|^2. \label{equ:a}
\end{equation*}
According to Lemma 2, we have 
\begin{equation*}
    \bar{\mathbf{x}}_{t+1} =\bar{\mathbf{x}}_{t}  - \bar{\mathbf{g}}_{t}, 
\end{equation*}
where $\bar{\mathbf{g}}_{t} = \frac{1}{N}\sum_{i=1}^N\mathbf{g}_t^{(i)}$.
Let $\overline{\nabla f}(\mathbf{x}_t) := \frac{1}{N} \sum_{i=1}^N \nabla f_i(\mathbf{x}_i)$. 
Taking the expectations of both sides, we have
\begin{align*}
   &\mathbb{E}f(\bar{\mathbf{x}}_{t+1}) - \mathbb{E}f(\bar{\mathbf{x}}_t) \\
   & \leq \langle \nabla f(\bar{\mathbf{x}}_t), \mathbb{E}\{\bar{\mathbf{x}}_{t+1} - \bar{\mathbf{x}}_t\}\rangle + \frac{L}{2} \mathbb{E}\| \bar{\mathbf{x}}_{t+1} - \bar{\mathbf{x}}_t\|^2 \\
   &= -\gamma\langle \nabla f(\bar{\mathbf{x}}_t), \mathbb{E}\bar{\mathbf{g}}_t\rangle + \frac{L}{2} \mathbb{E}\| \bar{\mathbf{x}}_{t+1} - \bar{\mathbf{x}}_t\|^2 \\
   &= -\gamma\langle \nabla f(\bar{\mathbf{x}}_t), \overline{\nabla f}(\mathbf{x}_t)\rangle + \frac{L}{2} \mathbb{E}\| \bar{\mathbf{x}}_{t+1} - \bar{\mathbf{x}}_t\|^2.
\end{align*}
For the first term, we have
\begin{align*}
    & \langle \nabla f(\bar{\mathbf{x}}_t), \overline{\nabla f}(\mathbf{x}_t)\rangle \\
    &= \frac{1}{2} (\| \nabla f(\bar{\mathbf{x}}_t)\|^2 + \|\overline{\nabla f}(\mathbf{x}_t)\|^2 -\|\nabla f(\bar{\mathbf{x}}_t) - \overline{\nabla f} (\mathbf{x}_t)\|^2),
\end{align*}
and $\|\nabla f(\bar{\mathbf{x}}_t) - \overline{\nabla f}(\mathbf{x}_t)\|^2$ can be bounded by
\begin{align*}
     \|\nabla f(\bar{\mathbf{x}}_t) - \overline{\nabla f}(\mathbf{x}_t)\|^2 &= \| \frac{1}{N} \sum_{i=1}^{N} [\nabla f_i(\bar{\mathbf{x}}_t) - \nabla f_i(\mathbf{x}_t^{(i)})]\|^2 \\
     &\leq \frac{1}{N} \sum_{i=1}^{N}\| \nabla f_i(\bar{\mathbf{x}}_t) - \nabla f_i(\mathbf{x}_t^{(i)})\|^2 \\ 
     &\leq \frac{L^2}{N} \sum_{i=1}^{N} \| \bar{\mathbf{x}}_t - \mathbf{x}_t^{(i)}\|^2.
\end{align*}
% where is due to Jensen's Inequality.
For the second term, we have
\begin{align*}
    & \frac{L}{2} \mathbb{E}\{\|\bar{\mathbf{x}}_{t+1} - \bar{\mathbf{x}}_t\|^2\}\\
    &=\frac{L}{2} \mathbb{E}\{\|\overline{\mathbf{x}}_{t+1} - \left(\overline{\mathbf{x}}_t - \gamma \overline{\mathbf{g}}_{t} \right) - \gamma \overline{\mathbf{g}}_{t} \|^2\} \\
    &= \underbrace{\frac{L}{2} \mathbb{E}\{\|\overline{\mathbf{x}}_{t+1} - \left(\overline{\mathbf{x}}_t - \gamma \overline{\mathbf{g}}_{t} \right)\|^2\}}_{T_1} + \underbrace{\frac{\gamma^2L}{2} \mathbb{E}\{\| \overline{\mathbf{g}}_{t} \|^2\}}_{T_2}. 
\end{align*}
%where \eqref{equ:split} is due to the randomness of communication network is independent with the random of stochastic gradients.
With Lemma 2, $T_1$ can be bounded by
\begin{align*}
    T_1 & \leq \frac{L\kappa}{2N^2} \sum_{i=1}^N \mathbb{E}\{\left(\|\mathbf{x}_i^{(t)} - \gamma \mathbf{g}_i^{(t)} - \bar{\mathbf{x}}_t + \gamma \bar{\mathbf{g}}_t\|^2 \right)\} \\
    & \leq \frac{L \kappa}{N^2}\sum_{i=1}^N \mathbb{E}\{\| \mathbf{x}_i^{(t)} -\bar{\mathbf{x}}_t\|^2\} + \frac{L \kappa \gamma^2}{N^2} \underbrace{\sum_{i=1}^N \mathbb{E}\{\| \mathbf{g}_i^{(t)}- \bar{\mathbf{g}}_t\|^2\}}_{T_3}.
\end{align*}
In addition, $T_2$ can be  bounded by
\begin{align*}
  & T_2  = \frac{\gamma^2L}{2} \mathbb{E}\{\|\frac{1}{N} \sum_{i=1}^N\left(\mathbf{g}_i^{(t)} - \nabla f_i(\mathbf{x}_i^{(t)}) + \nabla f_i(\mathbf{x}_i^{(t)}) \right)\|^2 \}\\
    &= \frac{\gamma^2L}{2N^2} \sum_{i=1}^N \mathbb{E}\{\|\mathbf{g}_i^{(t)} - \nabla f_i(\mathbf{x}_i^{(t)})\|^2\} \\
    & + \sum_{i=1}^N \mathbb{E}\{\frac{\gamma^2L}{2N^2}\| \nabla f_i(\mathbf{x}_i^{(t)})\|^2\}\\
    & \leq \frac{\gamma^2 L}{2N}\sigma^2 + \frac{\gamma^2L}{2} \mathbb{E}\{\| \overline{\nabla f}(\mathbf{x}_t)\|^2\}.
\end{align*}
$T_3$ is bounded by
\begin{align*}
    &T_3 = \sum_{i=1}^N  \mathbb{E}\{\|\left(\mathbf{g}_i^{(t)} - \nabla f_i(\mathbf{x}_i^{(t)})\right) + \left(\nabla f_i(\mathbf{x}_i^{(t)}) - \nabla f_i(\bar{\mathbf{x}}_t)\right) \\
    & + \left(\nabla f_i(\bar{\mathbf{x}}_t) - \nabla f(\bar{\mathbf{x}}_t)\right) - \left(\bar{\mathbf{g}}_t -\overline{\nabla f}(\mathbf{x}_t)\right) \\
    & - \left(\overline{\nabla f}(\mathbf{x}_t) - \nabla f(\bar{\mathbf{x}}_t)\right) \|^2 \}\\
    & = \sum_{i=1}^N \mathbb{E}\{ \|\left(\mathbf{g}_i^{(t)} - \nabla f_i(\mathbf{x}_i^{(t)}) - \bar{\mathbf{g}}_t + \overline{\nabla f}(\mathbf{x}_t)\right) \\
    & + \left(\nabla f_i(\mathbf{x}_i^{(t)}) - \nabla f_i(\bar{\mathbf{x}}_t)\right) + \left(\nabla f_i(\bar{\mathbf{x}}_t) - \nabla f(\bar{\mathbf{x}}_t)\right) \\
    & - \left(\overline{\nabla f}(\mathbf{x}_t) - \nabla f(\bar{\mathbf{x}}_t)\right)  \|^2 \}\\
    & = \sum_{i=1}^N  \mathbb{E}\{\|\frac{N-1}{N} \left(\mathbf{g}_i^{(t)} - \nabla f_i(\mathbf{x}_i^{(t)})\right)\\
    &  -\frac{1}{N} \sum_{j=1,j\neq i}^N \left(\mathbf{g}_j^{(t)} - \nabla f_j(\mathbf{x}_j^{(t)})\right) + \left(\nabla f_i(\mathbf{x}_i^{(t)}) - \nabla f_i(\bar{\mathbf{x}}_t)\right)\\
    & + \left(\nabla f_i(\bar{\mathbf{x}}_t) - \nabla f(\bar{\mathbf{x}}_t)\right) - \left(\overline{\nabla f}(\mathbf{x}_t) - \nabla f(\bar{\mathbf{x}}_t)\right)  \|^2 \}\\
    & = \sum_{i=1}^N  \mathbb{E}\{\|\frac{N-1}{N} \left(\mathbf{g}_i^{(t)} - \nabla f_i(\mathbf{x}_i^{(t)})\right)\|^2 \} \\
    & +  \sum_{i=1}^N \sum_{j=1,j\neq i}^N \mathbb{E}\{\| \frac{1}{N} \left(\mathbf{g}_j^{(t)} - \nabla f_j(\mathbf{x}_j^{(t)})\right)\|^2\} \\
    & + \sum_{i=1}^N \mathbb{E}\{\|\left(\nabla f_i(\mathbf{x}_i^{(t)}) - \nabla f_i(\bar{\mathbf{x}}_t)\right) + \left(\nabla f_i(\bar{\mathbf{x}}_t) - \nabla f(\bar{\mathbf{x}}_t)\right) \\
    & - \left(\overline{\nabla f}(\mathbf{x}_t) - \nabla f(\bar{\mathbf{x}}_t)\right)  \|^2 \}\\
    & \leq \frac{(N-1)^2}{N^2}  \sum_{i=1}^N \mathbb{E}\{\| \left(\mathbf{g}_i^{(t)} - \nabla f_i(\mathbf{x}_i^{(t)})\right) \|\}^2 \\
    & + \frac{1}{N} \sum_{i=1}^N \mathbb{E}\{ \| \left(\mathbf{g}_i^{(t)} - \nabla f_i(\mathbf{x}_i^{(t)})\right) \|^2 \} \\ 
    & + 3 \sum_{i=1}^N \mathbb{E}\{\|\left(\nabla f_i(\mathbf{x}_i^{(t)}) - \nabla f_i(\bar{\mathbf{x}}_t)\right)\|^2\}\\
    & + 3 \sum_{i=1}^N \mathbb{E}\{\|\left(\nabla f_i(\bar{\mathbf{x}}_t) - \nabla f(\bar{\mathbf{x}}_t)\right)\|^2\} \nonumber\\
    & + 3 \sum_{i=1}^N \mathbb{E}\{\|\left(\overline{\nabla f}(\mathbf{x}_t) - \nabla f(\bar{\mathbf{x}}_t)\right)\|^2\}\\
    & \leq \frac{N^2-2N+2}{N}\sigma^2 + 3NL^2 \sum_{i=1}^N \mathbb{E}\{ \|\mathbf{x}_i^{(t)} -\bar{\mathbf{x}}_t \|^2\} + 3N\zeta^2 \\
    & + 3L^2 \sum_{i=1}^N \mathbb{E}\{\|\mathbf{x}_i^{(t)} -\bar{\mathbf{x}}_t \|^2\} \\
    & \leq N\sigma^2 + 3(N+1)L^2 \sum_{i=1}^N  \mathbb{E}\{\|\mathbf{x}_i^{(t)} -\bar{\mathbf{x}}_t \|^2\} + 3N\zeta^2.  
\end{align*}
Putting everything back, we have 
\begin{align*}
   & \mathbb{E}\{f(\bar{\mathbf{x}}_{t+1})\} - \mathbb{E}\{f(\bar{\mathbf{x}}_t)\}  \\
   &\leq - \frac{\gamma}{2} \mathbb{E}\{\| \nabla f(\bar{\mathbf{x}}_t)\|^2\} - \frac{\gamma}{2} \mathbb{E}\{\| \overline{\nabla f}(\mathbf{x}_t)\|^2\} \\
   & + \frac{\gamma L^2}{2N} \sum_{i=1}^{N}\mathbb{E}\{\| \bar{\mathbf{x}}_t - \mathbf{x}_i^{(t)}\|^2 \} +\frac{\kappa L}{N^2} \sum_{i=1}^{N}\mathbb{E}\{\| \bar{\mathbf{x}}_t - \mathbf{x}_i^{(t)}\|^2\} \\
    &+ \frac{\kappa\gamma^2L}{N^2}(N\sigma^2 + (3N+1)L^2 \sum_{i=1}^{N}\mathbb{E}\{\| \bar{\mathbf{x}}_t - \mathbf{x}_t^{(i)}\|^2\} + 3N\zeta^2) \\
    & + \frac{\gamma^2 L}{2N}\sigma^2 + \frac{\gamma^2 L}{2}\sum_{i=1}^{N}\mathbb{E}\{\|\overline{\nabla f}(\bar{\mathbf{x}}_t) \|^2\} \\
    &= -\frac{\gamma}{2} \mathbb{E}\{\| \nabla f(\bar{\mathbf{x}}_t)\|^2\} -(\frac{\gamma}{2} - \frac{\gamma^2 L}{2}) \|\overline{\nabla f}(\mathbf{x}_t) \|^2 \\
    &+ (\frac{\gamma L^2}{2N} + \frac{L\kappa}{N^2} + \frac{(3N+1)L^3\gamma^2\kappa}{N^2})\sum_{i=1}^{N}\mathbb{E}\{\|\mathbf{x}_t^{(i)}- \bar{\mathbf{x}}_t \|^2\}\\
    &+ \frac{\gamma^2 L}{2N}\sigma^2 + \frac{\gamma^2 L\kappa}{N}\sigma^2 + \frac{3L\kappa\gamma^2\zeta^2}{N}.
\end{align*}
Summing over $t$ and taking the average, we can get
\begin{align*}
    & \frac{\mathbb{E}\{f(\bar{\mathbf{x}}_T) - f(\bar{\mathbf{x}}_0)\}}{T}\\
    & \leq -\frac{\gamma}{2T} \sum_{T=1}^T \mathbb{E}\{\| \nabla f(\bar{\mathbf{x}}_t)\|^2\} -(\frac{\gamma}{2T} - \frac{\gamma^2 L}{2T}) \sum_{T=1}^T \mathbb{E}\{\|\overline{\nabla f}(\mathbf{x}_t) \|^2\} \\
    &+ (\frac{\gamma L^2}{2N} + \frac{L\kappa}{N^2} + \frac{(3N+1)L^3\gamma^2\kappa}{N^2})\frac{1}{T} \sum_{t=1}^T\sum_{i=1}^{N}\mathbb{E}\{\|\mathbf{x}_t^{(i)}- \bar{\mathbf{x}}_t\|^2\}\\
    &+ \frac{\gamma^2 L}{2N}\sigma^2 + \frac{\gamma^2 L\kappa}{n}\sigma^2 + \frac{3L\kappa\gamma^2\zeta^2}{N}.
\end{align*}
By minor rearranging, we get
\begin{align*}
   & \frac{1}{T} \sum_{T=1}^T  \| \nabla f(\bar{\mathbf{x}}_t)\|^2 \\
   &\leq \frac{2\mathbb{E}[f(\bar{\mathbf{x}}_0)] - \mathbb{E}[f(\bar{\mathbf{x}}_T)]}{\gamma T} - \frac{1-\gamma L}{T}  \sum_{T=1}^T  \mathbb{E}\{\|\overline{\nabla f}(\mathbf{x}_t) \|^2\}  \\
   & + (\frac{L^2}{NT} + \frac{2 L\kappa}{\gamma N^2T} + \frac{2(3N+1)L^3\gamma\kappa}{N^2T}) \sum_{t=1}^T \sum_{i=1}^{N} \mathbb{E}\{\| \mathbf{x}_t^{(i)}- \bar{\mathbf{x}}_t \|^2\}\nonumber\\
   & + \frac{\gamma L}{N}\sigma^2 + \frac{2 \gamma L\kappa}{N}\sigma^2 + \frac{6L\kappa\gamma\zeta^2}{N}.
\end{align*}

% ==== Now bounding Variance ======== 
Next we are going to bound $\sum_{t=1}^T \sum_{i=1}^{N} \mathbb{E}\{\| \mathbf{x}_t^{(i)}- \bar{\mathbf{x}}_t \|^2\}$.
Letting $\bm{\beta}(t) = (\mathbf{I} - \mathbf{J})\mathbf{X}_t$, we have $\|\bm{\beta}(t)\|_F^2 = \sum_{i=1}^N \|\mathbf{x}_i^{(t)} - \overline{\mathbf{x}}_t \|^2$.

Let's consider the $l$-th dimension. From Lemma 3, we have
\begin{align*}
    \bm{\beta}_l(t)&= (\widetilde{\mathbf{W}}_l^{(t-1)} - \mathbf{J}) \left( \mathbf{X}_l^{(t-1)}  - \gamma \mathbf{G}_l^{(t-1)}  \right) \\
              &= \widetilde{\mathbf{W}}_l^{(t)}  \bm{\beta}_l(t-1) - \gamma (\widetilde{\mathbf{W}_l}^{(t)} - \mathbf{J}) \mathbf{G}_l^{(t-1)} \\
              &= \cdot\cdot\cdot \\
              &= \prod_{k=1}^{t-1} \widetilde{\mathbf{W}}_l^{(t)}\bm{\beta}_l(0) -\gamma \sum_{k=1}^{t-1} \left( (\prod_{m=1}^{t-1}  \widetilde{\mathbf{W}}_l^{(t)} - \mathbf{J}) \right) \mathbf{G}_l^{(k)}.
\end{align*}
Since all devices are initialized with $\mathbf{x}_0$, $\bm{\beta}_l(0) = 0$, we have, 
\begin{align*}
    & \mathbb{E}\{\|\bm{\beta}(t)\|_F^2\} = \sum_{l=1}^d \mathbb{E}\{\|\bm{\beta}_l(t)\|^2\}\\
    & = \gamma^2 \sum_{l=1}^d \mathbb{E}  \| \sum_{k=1}^{t-1} \left( (\prod_{m=1}^{t-1}  \widetilde{\mathbf{W}}_l^{(t)} - \mathbf{J}) \right) \mathbf{G}_l^{(k)} \|^2  \\
             %& = \gamma^2 \sum_{l=1}^d \mathbb{E}  \| \sum_{k=1}^{t-1} \left( (\prod_{m=1}^{t-1}  \widetilde{\mathbf{W}}_l^{(t)} - \mathbf{J}) \right) (\mathbf{G}_l^{(k)} - \nabla \mathbf{F}_l^{(k)} + \nabla \mathbf{F}_l^{(k)}) \|^2  \\
             & = 2\gamma^2 \underbrace{\sum_{l=1}^d \mathbb{E}  \| \sum_{k=1}^{t-1} \left( (\prod_{m=1}^{t-1}  \widetilde{\mathbf{W}}_l^{(t)} - \mathbf{J}) \right) (\mathbf{G}_l^{(k)} - \nabla \mathbf{F}_l^{(k)}) \|^2}_{T_4} \nonumber \nonumber\\
             & +  2\gamma^2 \underbrace{ \sum_{l=1}^d \mathbb{E} \| \sum_{k=1}^{t-1} \left( (\prod_{m=1}^{t-1}  \widetilde{\mathbf{W}}_l^{(t)} - \mathbf{J}) \right) (\nabla \mathbf{F}_l^{(k)})}_{T_5} \|^2.
\end{align*}
For $T_4$, we have
\begin{align*}
    \mathbb{E} \{T_4\} &= \sum_{l=1}^d \sum_{k=1}^{t-1} \mathbb{E} \{\|\left( (\prod_{m=1}^{t-1}  \widetilde{\mathbf{W}}_l^{(t)} - \mathbf{J}) \right) (\mathbf{G}_l^{(k)} - \nabla \mathbf{F}_l^{(k)}) \|^2 \} \\
      & \leq\sum_{l=1}^d \sum_{k=1}^{t-1} \rho^{t-k}  \mathbb{E}\{ \|\mathbf{G}_l^{(k)} - \nabla \mathbf{F}_l^{(k)}\|^2\} \\
      & \leq \sum_{k=1}^{t-1} \rho^{t-k}  \mathbb{E}\{ \|\mathbf{G}_k - \nabla \mathbf{F}_k\|_F^2\}   \\
      & \leq N \sigma^2 \rho(1 + \rho + \rho^2 + \cdot \cdot \cdot + \rho^t) \\
      & \leq \frac{N\sigma^2\rho}{1-\rho}.
\end{align*}
For the second term, let $\mathbf{A}_{q,p} := \prod_{m=q}^p \widetilde{\mathbf{W}}_l^{(t)} - \mathbf{J}$, then
\begin{align*}
    & \mathbb{E} \{T_5\} = \sum_{l=1}^d  \sum_{k=1}^{t-1} \mathbb{E} \{ \| \mathbf{A}_{k,t-1} \nabla \mathbf{F}_l^{(k)} \|^2 \} \nonumber\\
    & + \sum_{l=1}^d \sum_{k=1}^{t-1} \sum_{m=1, m\neq k}^{t-1}\mathbb{E} \{ (\mathbf{A}_{k,t-1} \nabla \mathbf{F}_l^{(k)})^T (\mathbf{A}_{m,t-1} \nabla \mathbf{F}_l^{(m)}) \} \\
    &\leq  \sum_{l=1}^d  \sum_{k=1}^{t-1} \rho^{t-k}\mathbb{E} \{ \| \nabla \mathbf{F}_l^{(k)} \|^2 \nonumber\\ 
    &+ \sum_{l=1}^d \sum_{k=1}^{t-1} \sum_{m=1, m\neq k}^{t-1}\mathbb{E} \{ \|\mathbf{A}_{k,t-1} \nabla \mathbf{F}_l^{(k)})\| \|(\mathbf{A}_{m,t-1} \nabla \mathbf{F}_l^{(m)})\| \}\\
    & \leq  \sum_{k=1}^{t-1} \rho^{t-k} \mathbb{E}\{ \| \nabla \mathbf{F}_k \|_F^2 \nonumber\\
    & + \sum_{l=1}^d \sum_{k=1}^{t-1} \sum_{m=1, m\neq k}^{t-1}\mathbb{E} \{ \frac{1}{2\epsilon}\|\mathbf{A}_{k,t-1} \nabla \mathbf{F}_l^{(k)})\|^2+ \frac{\epsilon}{2}\|(\mathbf{A}_{m,t-1} \nabla \mathbf{F}_l^{(m)})\|^2 \}\\
    & \leq  \sum_{k=1}^{t-1} \rho^{t-k} \mathbb{E}\{ \| \nabla \mathbf{F}_k \|_F^2  \\
    & + \sum_{l=1}^d \sum_{k=1}^{t-1} \sum_{m=1, m\neq k}^{t-1}\mathbb{E} \{ \frac{\rho^{t-k}}{2\epsilon}\|\nabla  \mathbf{F}_l^{(k)})\|^2+ \frac{\epsilon \rho^{t-m}}{2}\| \nabla \mathbf{F}_l^{(m)})\|^2 \} \\
    & = \sum_{k=1}^{t-1} \rho^{t-k} \mathbb{E}\{ \| \nabla \mathbf{F}_k \|_F^2 \\
    & + \sum_{k=1}^{t-1} \sum_{m=1, m\neq k}^{t-1}\mathbb{E} \{ \frac{\rho^{t-k}}{2\epsilon}\|\nabla \mathbf{F}_k\|_F^2 + \frac{\epsilon\rho^{t-m}}{2}\|\mathbf{F}_m\|_F^2.
\end{align*}
%where \eqref{equ:young} is due to Young's Inequallity: $2ab \leq \frac{a^2}{\epsilon} + \epsilon b^2$. 
By setting $\epsilon = \rho^{\frac{p-q}{2}}$, we have
\begin{align*}
    & \mathbb{E}\{T_5\} \leq \sum_{k=1}^{t-1} \rho^{t-k} \mathbb{E}\{ \| \nabla \mathbf{F}_k \|_F^2\}  \\
    & +  \frac{1}{2} \sum_{k=1}^{t-1} \sum_{m=1, m\neq k}^{t-1}\mathbb{E} \{ \sqrt{\rho}^{2t-k-m} \left( \| \nabla \mathbf{F}_k \|_F^2 + \| \nabla \mathbf{F}_m \|^2\right) \\
    & \leq \sum_{k=1}^{t-1} \rho^{t-k} \mathbb{E}\{ \| \nabla \mathbf{F}_k \|_F^2\} \\
    & + \sum_{k=1}^{t-1} \left( \sqrt{\rho}^{t-k} \mathbb{E}\{ \|\mathbf{F}_k\|_F^2\} \cdot \sum_{m=1, m\neq k}^{t-1} \sqrt{\rho}^{t-m} \right)\\
    & \leq \sum_{k=1}^{t-1} \rho^{t-k} \mathbb{E}\{ \| \nabla \mathbf{F}_k \|_F^2 \} \\
    & + \sum_{k=1}^{t-1} \left( \sqrt{\rho}^{t-k} \mathbb{E}\{ \|\mathbf{F}_k\|_F^2\} \cdot \sum_{m=1}^{t-1} \sqrt{\rho}^{t-m} - \sqrt{\rho}^{t-k}\right) \\
    & \leq \frac{\sqrt{\rho}}{1-\sqrt{\rho}} \sum_{k=1}^{t-1}\sqrt{\rho}^{t-k} \mathbb{E} \| \nabla \mathbf{F}_k\|_F^2.
\end{align*}
Putting $T_4$ and $T_5$ back, we have 
\begin{align*}
    &\frac{1}{NT} \sum_{t=1}^T \mathbb{E}\{\|\bm{\beta(t)}\|\} \\
    &\leq \frac{2 \gamma^2 \sigma^2 \rho}{1-\rho} + \frac{2 \gamma^2 \sqrt{\rho}}{(1-\sqrt{\rho})NT} \sum_{t=1}^T \sum_{k=1}^{t-1} \sqrt{\rho}^{t-k} \mathbb{E} \| \nabla \mathbf{F}_k\|^2\\
    & = \frac{2 \gamma^2 \sigma^2 \rho}{1-\rho} + \frac{2 \gamma^2 \sqrt{\rho}}{(1-\sqrt{\rho})NT} \sum_{t=1}^T \mathbb{E}\{\| \nabla \mathbf{F}_t\|_F^2\} \sum_{k=1}^{T-t}\sqrt{\rho}^k \\
    & \leq \frac{2 \gamma^2 \sigma^2 \rho}{1-\rho} + \frac{2 \gamma^2 \sqrt{\rho}}{(1-\sqrt{\rho})NT} \sum_{t=1}^T \mathbb{E}\{\| \nabla \mathbf{F}_t\|^2\} \frac{\sqrt{\rho}}{1-\sqrt{\rho}} \\
    & =  \frac{2 \gamma^2 \sigma^2 \rho}{1-\rho} + \frac{2 \gamma^2 \rho}{(1-\sqrt{\rho})^2NT} \sum_{t=1}^T \mathbb{E}\{\| \nabla \mathbf{F}_t\|^2\}.
\end{align*}
In addition, $\| \nabla \mathbf{F}_t\|_F^2$ can be bounded by
\begin{align*}
    &\| \nabla \mathbf{F}_t\|_F^2 = \sum_{i=1}^N \| \nabla f(\mathbf{x}_i^{(t)}) \|^2 \\
    &= \sum_{i=1}^N \| \nabla f_i(\mathbf{x}_i^{(t)}) - \nabla f(\mathbf{x}_i^{(t)}) + \nabla f(\mathbf{x}_i^{(t)})\\
    &- \nabla f(\bar{\mathbf{x}_t}) + \nabla f(\bar{\mathbf{x}_t})\|^2 \\
    & \leq 3 \sum_{i=1}^N \|\nabla f_i(\mathbf{x}_i^{(t)}) - \nabla f(\mathbf{x}_i^{(t)}) \|^2 \\
    & + 3 \sum_{i=1}^N \|\nabla f(\mathbf{x}_i^{(t)}) - \nabla f(\bar{\mathbf{x}}_t) \|^2 \\
    & + 3 \sum_{i=1}^N \|\nabla f(\bar{\mathbf{x}}_t) \|^2 \\
    & \leq 3N\zeta^2 + 3L^2 \sum_{i=1}^N \|\mathbf{x}_t^{(i)}- \bar{\mathbf{x}}_t \|^2 + 3N \| \nabla f(\bar{\mathbf{x}}_t)\|^2. 
\end{align*}
Plugging back we have
\begin{align*}
    &\frac{1}{NT}\sum_{t=1}^T \sum_{i=1}^{N}\mathbb{E}\{\| \mathbf{x}_t^{(i)}- \bar{\mathbf{x}}_t \|^2\} \\
    &\leq   \frac{6N\gamma^2L^2 \rho}{(1-\sqrt{\rho})NT} \sum_{t=1}^T \sum_{i=1}^N\mathbb{E}\{\| \mathbf{x}_t^{(i)}- \bar{\mathbf{x}}_t\|^2 \}\\
    & + \frac{6N\gamma^2L^2 \rho}{(1-\sqrt{\rho})NT} \sum_{t=1}^T \mathbb{E} \| \nabla f(\overline{\mathbf{x}}_t)\|^2 \\
    & + \frac{2 \gamma^2 \sigma^2 \rho}{1-\rho} + \frac{6\zeta^2\gamma^2 \sqrt{\rho}}{(1-\sqrt{\rho})}.
\end{align*}
After minor rearrangement, we have
\begin{align*}
    & \sum_{t=1}^T \sum_{i=1}^N\mathbb{E}\{\| \mathbf{x}_t^{(i)}- \bar{\mathbf{x}}_t\|^2 \} \\
    & \leq \frac{NT}{1-D}(\frac{2\gamma^2\sigma^2\rho}{1-\rho} + \frac{6\gamma^2\zeta^2\rho}{(1-\sqrt{\rho})^2}\\
    & + \frac{6\gamma^2\rho}{(1-\sqrt{\rho})^2 T} \sum_{t=1}^T  \mathbb{E}\{\| \nabla f(\bar{\mathbf{x}}_t)\|^2\}).
\end{align*}
where $D = \frac{6\gamma^2 L^2\rho}{(1-\sqrt{\rho})^2}$.
Plugging back to original, we have
\begin{align*}
     &\frac{1}{T}\sum_T \mathbb{E}\{\| \nabla f(\bar{\mathbf{x}}_t)\|^2 \} \\
     &\leq  \frac{\mathbb{E}[f(\bar{\mathbf{x}}_0)] - \mathbb{E}[f(\bar{\mathbf{x}}_{T})]}{\gamma T} - \frac{1-\gamma L}{T}  \sum_{t=1}^T \mathbb{E}\{\|\overline{\nabla f}(\bar{\mathbf{x}}_t) \|^2 \}\\
     & + (\frac{L^2}{NT} + \frac{2 L\kappa}{\gamma N^2T} + \frac{2(3N+1)L^3\gamma\kappa}{N^2T})\frac{NT}{(1-D)}\\
     & (\frac{2\gamma^2\sigma^2\rho}{1-\rho} + \frac{6\gamma^2\zeta^2\rho}{(1-\sqrt{\rho})^2} +\frac{6\gamma^2\rho}{(1-\sqrt{\rho})^2 T} \sum_{t=1}^T \mathbb{E}\{\| \nabla f(\bar{\mathbf{x}}_t)\|^2\}) \\
     & + \frac{\gamma L}{N}\sigma^2 + \frac{2 \gamma L\kappa}{N}\sigma^2 + \frac{6L\kappa\gamma \zeta^2}{N}
\end{align*}
Therefore,
\begin{align*}
     & \frac{1}{T}\sum_{t=1}^T \mathbb{E}\{\| \nabla f(\bar{\mathbf{x}}_t)\|^2 \} \\
     &\leq (\frac{\mathbb{E}[f(\bar{\mathbf{x}}_0)] - \mathbb{E}[f(\bar{\mathbf{x}}_{T})]}{\gamma T} + \frac{\gamma L}{N}\sigma^2 + \frac{2 \gamma L\kappa}{N}\sigma^2 + \frac{6L\kappa\gamma \zeta^2}{N}) \frac{1-D}{1-2D} \nonumber\\
     &+ (L^2 + \frac{2 L\kappa}{\gamma N} + \frac{2(3N+1)L^3\gamma\kappa}{N})(\frac{2\gamma^2\sigma^2\rho}{1-\rho}  +\frac{6\gamma^2\zeta^2\rho}{(1-\sqrt{\rho})^2}) \frac{1}{1-2D}.
\end{align*}
Recall that we require that $\gamma L \leq \frac{(1-\sqrt{\rho})}{4\sqrt{\rho}}$. Therefore,
\begin{equation*}
    D = \frac{6\gamma^2 L^2 \rho}{(1-\sqrt{\rho})^2} \leq \frac{3}{8}. 
\end{equation*}
Plugging and setting $\gamma=\sqrt{\frac{N}{T}}$, we have
\begin{align*}
    &\frac{1}{T}\sum_{t=1}^T \mathbb{E}\{\| \nabla f(\bar{\mathbf{x}}_t)\|^2 \}\\
    &\leq \frac{8}{\sqrt{NT}}(\mathbb{E}[f(\bar{\mathbf{x}}_0)] - \mathbb{E}[f(\bar{\mathbf{x}}_{T})] + L\sigma^2 + 2L\kappa\sigma^2\\
    & + 6L\kappa \zeta^2 + 2L\kappa (\frac{2\sigma^2\rho}{1-\rho}  +\frac{6\zeta^2\rho}{(1-\sqrt{\rho})^2})) \\
    & +\frac{8N}{T} (L^2 + \frac{2(3N+1)L^3\kappa}{\sqrt{TN}})(\frac{2\sigma^2\rho}{1-\rho}  +\frac{6\zeta^2\rho}{(1-\sqrt{\rho})^2})\\
    &= \mathcal{O}(\frac{1}{\sqrt{NT}}) + \mathcal{O}(\frac{N}{T}).
\end{align*}
\end{proof}
%% ======== Prove of the convexity ====

\subsection{Proof of Convexity of Weight Optimization Problem in (13)}
\begin{proof} Let $\mathcal{S} =\{ \mathbf{W}| 0 \leq w_{i, j} \leq 1, \mathbf{W} = \mathbf{W}^T , \mathbf{W} \mathbf{1} =  \mathbf{1}\}$.
For $\mathbf{W}_A$ and $\mathbf{W} _B$ from $\mathcal{S}$,  and letting $\mathbf{W}_C = \eta \mathbf{W}_A + (1-\eta) \mathbf{W}_B$, where $0 \leq \eta \leq 1$, it is easy to verify that $\mathbf{W}_C \in \mathcal{S}$.
As shown in Lemma 1, $\overline{\mathbf{W}^2} = \mathbb{E}\{ \widetilde{\mathbf{W}}^T \widetilde{\mathbf{W}}\}$, where $\widetilde{\mathbf{W}} = \mathbf{W} \odot \mathbf{A}  + \mathbf{I} - \text{Diag}(\mathbf{W} \mathbf{A})$ and $\mathbf{A}$ represents the successfulness of transmission.
For a given $\mathbf{A}$, we have $\widetilde{\mathbf{W}}_C = \eta \widetilde{\mathbf{W}}_A + (1-\eta) \widetilde{\mathbf{W}}_B$.

Hence, we can get
\begin{align*}
    \widetilde{\mathbf{W}}_C^T \widetilde{\mathbf{W}}_C &= \left(\eta \widetilde{\mathbf{W}}_A + (1-\eta) \widetilde{\mathbf{W}}_B\right)^T\left(\eta \widetilde{\mathbf{W}}_A + (1-\eta) \widetilde{\mathbf{W}}_B\right) \\
    & = \eta^2 \widetilde{\mathbf{W_A}}^T \widetilde{\mathbf{W}}_A + (1-\eta)^2 \widetilde{\mathbf{W}}_B^T \widetilde{\mathbf{W}}_B \\
    & + \eta(1-\eta) \widetilde{\mathbf{W}}_A^T \widetilde{\mathbf{W}}_B + \eta(1-\eta) \widetilde{\mathbf{W}}_B^T \widetilde{\mathbf{W}} _A.
\end{align*}
We also have
\begin{align*}
& \widetilde{\mathbf{W}}_C^T \widetilde{\mathbf{W}}_C - \left( \eta \widetilde{\mathbf{W}}_A^T \widetilde{\mathbf{W}}_A + (1-\eta) \widetilde{\mathbf{W}}_B^T \widetilde{\mathbf{W}}_B \right) \\
&= -\eta(1-\eta)\left(\widetilde{\mathbf{W}}_A-\widetilde{\mathbf{W}}_B\right)^T \left(\widetilde{\mathbf{W}}_A-\widetilde{\mathbf{W}}_B\right) \preceq \mathbf{0}.
\end{align*}

This implies
\begin{equation*}
    \widetilde{\mathbf{W}}_C^T \widetilde{\mathbf{W}}_C  \preceq \eta \widetilde{\mathbf{W}}_A^T \widetilde{\mathbf{W}}_A + (1-\eta) \widetilde{\mathbf{W}}_B^T \widetilde{\mathbf{W}}_B.
\end{equation*}
Considering the objective, we can get
\begin{align*}
    \widetilde{\mathbf{W}}_C^T \widetilde{\mathbf{W}}_C - \mathbf{J} \preceq & \eta \left(\widetilde{\mathbf{W}}_A^T \widetilde{\mathbf{W}}_A -\mathbf{J} \right) \\
    & + (1-\eta) \left(\widetilde{\mathbf{W}}_B^T \widetilde{\mathbf{W}}_B - \mathbf{J}\right)
\end{align*}
Taking the expectation of both sides, we have
\begin{align*}
    \overline{\mathbf{W_C}^2} - \mathbf{J} \preceq & \eta \left(\overline{\mathbf{W_A}^2} -\frac{1}{N} \mathbf{11^T} \right)\\
     & + (1-\eta) \left(\overline{\mathbf{W_B}^2}- \mathbf{J}\right)
\end{align*}
Since $\lambda_{max}$ is a convex function, we have
\begin{align*}
    \lambda_{max} \{\overline{\mathbf{W_C}^2} - \mathbf{J}\} \leq & \lambda_{max} (\eta \left(\overline{\mathbf{W_A}^2} -\mathbf{J} \right) \\
    & + (1-\eta) \left(\overline{\mathbf{W_B}^2}- \mathbf{J}\right)) \\
    \leq & \eta \lambda_{max} \left(\overline{\mathbf{W_A}^2} -\mathbf{J} \right) \\
    & + (1-\eta) \lambda_{max} \left(\overline{\mathbf{W_B}^2}- \mathbf{J}\right)
\end{align*}
Hence, the objective of the weight optimization problem is convex.
\end{proof}

\bibliographystyle{IEEEtran}
\bibliography{main}
\end{document}